\documentclass[onecolumn,english,preprintnumbers,showkeys,nofootinbib,pra]{revtex4}

\usepackage{amsmath,amsfonts,amssymb}
\usepackage{graphicx}
\usepackage[colorlinks=true, allcolors=blue]{hyperref}

\usepackage{bm}
\usepackage{mathrsfs}
\usepackage{feynmf}
\usepackage{slashed}
\usepackage{verbatim}

\begin{document}

\title{Quantum lattice gas  algorithmic representation \\of gauge field theory}

\author{Jeffrey Yepez}\email{yepez@hawaii.edu}
\date{September 14, 2016}
 \affiliation{
  Department\;of\;Physics\;and\;Astronomy,  University\;of\;Hawai`i\;at\;M\=anoa, Watanabe\;Hall,\;2505\;Correa\;Road, Honolulu,\;Hawai`i\;96822
}

\begin{abstract}
Presented is a quantum lattice gas algorithm to efficiently model a system of Dirac particles interacting through an intermediary gauge field. The  algorithm uses a fixed qubit array to represent both the spacetime and the particles contained in the spacetime. Despite being a lattice based algorithm, Lorentz invariance is preserved down to the grid scale, with the continuum Dirac Hamiltonian generating the local unitary evolution even at  that scale: there is nonlinear scaling between the smallest observable time and that time measured in the quantum field theory limit, a kind of time dilation effect that emerges on small scales but has no effect on large scales. The quantum lattice gas algorithm correctly accounts for the anticommutative braiding of indistinguishable fermions---it does not suffer the Fermi-sign problem. It provides a highly convergent numerical simulation for strongly-correlated fermions  equal to a covariant path integral, presented here for the case when a Dirac particle's Compton wavelength is large compared to the grid scale of the qubit array.
\end{abstract}

\keywords{Quantum lattice gas, quantum computing, many-fermion quantum simulation, path summation, gauge field theory, quantum electrodynamics, superconducting fluid, Dirac-Maxwell-London equations}

\maketitle

\section{Introduction}
\label{sec:intro}

 Quantum computing for quantum simulation  \cite{chuang_Nature98} has  emerged as an active area at the nexus of quantum information science and 
  quantum physics.  Analog quantum simulations  have  demonstrated quantum phenomena theoretically predicted long ago but not experimentally accessible until recently,  for example 
relativistic effects of zitterbewegung \cite{1367-2630-15-7-073011} and Klein's tunneling paradox in the Dirac equation \cite{PhysRevLett.98.253005,4734301720100107,PhysRevLett.106.060503}.  In addition to experiments,  quantum simulation methods can be tested on conventional digital electronic computers and supercomputers.  A number of quantum simulation methods have been proposed  as practical computational physics methods and tested on conventional computers \cite{bialynicki-birula-prd94,PhysRevA.59.604,PhysRevA.86.062105,Zalka:1996p1167,succi-PhysRevE96,succi-ijmpc98,succi-cpc02,Zalka_PRSL_98}.  
The quantum simulation method  presented here is based on the quantum lattice gas model  of quantum computation. Quantum lattice gas models are intended for implementation on a Feynman quantum computer \cite{feynman-ces60,feynman-82} constructed out of quantum gates  \cite{feynman-85,divincenzo-pra95,divincenzo-pra95a,barenco-prsl95},  and thus a quantum lattice gas can be architected as a quantum circuit network. 
The quantum lattice gas model presented  here is a member of a class of unitary quantum algorithms that can serve as a testbed for exploring superfluid  dynamics  \cite{yepez:084501}.\footnote{Coherent quantum matter in the zero-temperature limit behaves like a  nonlinear macroscopic quantum particle, a nonlinear quantum system that is readily modeled on a supercomputer.}
   Quantum lattice gas models for many-body quantum simulations have  been investigated for nonrelativistic dynamics \cite{boghosian-ijmp97,boghosian-pd98,boghosian-pre98,PhysRevE.57.54}, relativistic dynamics \cite{yepez-qip-05,yepez_arXiv1106.0739_gr_qc,yepez_arXiv1512.02550_quant_ph}, and gravitational dynamics in the weak-gravity limit \cite{yepez:770202}.   Here  a quantum lattice gas simulation model  is introduced for  quantum simulation of  quantum field theories.  Two  example applications are given for gauge field theories   possessing an Abelian gauge group:   (1) superconducting  fluid dynamics and  (2) quantum electrodynamics.

The quantum lattice gas method has the distinguishing feature that it is constructed using a particle-based metaphor whereby quantum computation is reduced to the dynamical motion and interaction of bits encoded in an array of quantum bits---herein referred to as a qubit array. It has other salient features as well.  A quantum  engineering-related feature is that a quantum lattice gas bridges the gap between quantum simulation carried out on an analog quantum computer and  quantum simulation carried out on a quantum-gate-based Feynman quantum computer.   A quantum physics-related feature is that a quantum lattice gas can serve as a theoretical technique for investigating  many-fermion dynamics,  and this technique is particularly useful for understanding the behavior of a system of Fermi particles interacting via a spacetime-dependent intermediary gauge field.  It models the interactions as a strictly local unitary process in a finite size Hilbert space and therefore is an exactly computable representation.    The algorithmic protocol  builds directly upon the unitary update  stream and collide rule of  the previous quantum lattice gas algorithm for the Dirac equation \cite{yepez-qip-05,yepez_arXiv1307.3595_quant_ph,yepez_arXiv1106.0739_gr_qc,yepez_arXiv1512.02550_quant_ph},  applying  that quantum algorithm to the fermionic matter field and generalizing it for the Maxwell equations describing the bosonic intermediary gauge field.  The key to developing a quantum lattice gas algorithm for the coupled set of Dirac and Maxwell equations is to avoid trying to directly model the Maxwell equations by themselves.  In the approach presented here, the coupled Dirac-Maxwell equations are recovered as a limit of a more general set of coupled equations for a superconducting quantum fluid comprised of strongly-correlated fermions---the coupled Dirac-Maxwell-London equations.  
 
 The  Dirac-Maxwell-London equations are  a coupled set of equations of motion for a 4-spinor Dirac field, a massive 4-potential field, a second-rank  electromagnetic field tensor, and a 4-current source field.  The quantum lattice gas representation introduced here uses an equivalent and novel set of coupled equations expressed concisely with the 4-spinor Dirac field and a pair of Majorana-like 4-spinor fields.  The desired Dirac-Maxwell equations of quantum electrodynamics are recovered in the limit where the mass of the 4-potential field approaches zero (London penetration depth approaches the size of the system).    
The spinor representation of the Maxwell equations of electrodynamics was originally discovered by Laporte and Uhlenbeck just a few years following the discovery of the Dirac equation \cite{PhysRev.37.1380} and a year before Majorana's paper on representing spin-$f$ bosons by $2f$ fermions \cite{Majorana_Nuovo_Cimento_1932}. Laporte, Uhlenbeck, and Majorana-like representations of quantum fields is foundational to the quantum lattice gas model presented here.  Furthermore,  Bialynicki-Birula presented the first lattice model of both the Dirac equation and the spinor form of the Maxwell equations \cite{PhysRevD.49.6920}, closely following  Feynman's approach of particle dynamics confined to a spacetime lattice \cite{feynman-cit46,feynman-65-1st-qlga}, an approach continued by Jacobson \cite{jacobson-jpamg84,jacobson-jpamg.v17.84}. 

 Feynman's path integral approach, carried out on a spacetime lattice, is also foundational to the quantum lattice gas model presented here.  Feynman's concept of a spacetime lattice is generalized to be a qubit array.  The quantum dynamics of many-fermion system with gauge field interactions is expressed as a unitary path summation on the qubit array that is congruent to a relativistic path integral based on a covariant Lagrangian density of the modeled quantum field theory. The qubit-encoded spacetime lattice  acts as a regulator, whereby the quantum lattice gas representation of quantum field theory avoids infinities in calculated quantities.  Furthermore, renormalization is not needed to compensate for effects  due to self-interactions.  So the quantum lattice gas model can provide accurate numerical predictions when  it is employed as an effective field theory in the intended scale where the fermion's Compton wavelength is much larger than the smallest grid scale of the qubit array.  The quantum lattice gas method is a computational physics method for  high-energy physics applications  in gauge field theories requiring time-dependent analysis and also for  low-temperature physics applications requiring analysis of nonequilibrium effects in superconducting fluids.

\vspace{-1em}
  \subsection{Organization}
  
Presented is a quantum lattice gas model that can serve as a discrete representation of the  
Dirac-Maxwell-London equations for a superconducting fluid and, in the limit of vanishing London mass, that can also serve as a discrete representation of the Dirac-Maxwell equations of quantum electrodynamics.   Sec.~\ref{Section_Gauge_field_theory} presents two examples of gauge field theory.  The first example is quantum electrodynamics for matter field $\psi$ interacting via a 4-potential field $A^\mu$ driven with a source 4-current $e J^\mu$, and the  second example is a superconducting quantum fluid where the 4-potential is proportional to the 4-current,  $A^\mu = - \lambda_\text{\tiny L}^2 e J^\mu$, where $\lambda_\text{\tiny L}$ is the London penetration depth and $e$ is the electric charge of the Dirac particle.  The equations of motion are presented using covariant 4-vector notation as well as using  4-spinor notation.
Sec.~\ref{Section_quantum_lattice_gas_model} introduces the quantum lattice gas method.  The equations of motion for a superconducting quantum fluid (the Dirac-Maxwell-London equations) are cast in a discrete space representation, which can be in turn directly written in manifestly unitary form. The unitary form of the equations of motion is the basis of a local update rule using a stream and collide protocol that is provided at the end of the section. 
Sec.~\ref{QEQ_quantum_algorithm} presents the quantum lattice gas algorithm for a superconducting quantum fluid, and for quantum electrodynamics as a special case in the limit where $\lambda_\text{\tiny L}$ approaches the size of the system (London mass $m_\text{\tiny L}$ approaches the highest energy scale $\hbar/\tau$ in lattice units with $c=1$).  A unitary path summation on the qubit array based on a quantum lattice gas Hamiltonian operator is shown to be equivalent a path integral based on a covariant Lagangian density functional. 
Sec.~\ref{Section_Conclusion} concludes with some final remarks about  Feynman's quantum computing conjecture.  This section reviews why the quantum lattice gas algorithm is expected to avoid the Fermi sign problem, and it closes with some future outlooks.

Mathematical expository material is relegated to a number of appendices to  make the main presentation more accessible.  
Appendix~\ref{A_mu_field_dynamics} gives a derivation of the  4-spinor representation of the Maxwell equations.  This is needed to construct a quantum algorithm for the Maxwell equations. 
Appendix~\ref{Sec_derivation_Maxwell_London_equations_Yepez_form} gives a derivation of the  4-spinor representation of the Maxwell-London equations. 
Appendix~\ref{Sec_Bloch_Wannier_continuous_field_picture},  presents the Bloch-Wannier picture used to describe quantum particle dynamics in solid-state systems.  Delocalized Bloch wave and local Wannier states are continuous wave packet descriptions of particle dynamics that follow from representing a confining lattice as an external periodic potential.  The opportunity to switch from the lattice picture with a discrete set of points to the Bloch-Wannier picture with a continuous space of points allows one to switch the quantum lattice equation from discrete and finite-state variables to continuous variables. 
Appendix~\ref{Sec_derivation_Maxwell_London_equations_Yepez_generalized_Dirac_form} demonstrates that the  4-spinor representation of the Maxwell-London equations can be cast as a generalized Dirac equation for an 8-component field, comprised of a pair of  4-spinors.  This  generalized Dirac equation is needed to construct a quantum algorithm for the Maxwell-London equations.

\section{Gauge field theory}
\label{Section_Gauge_field_theory}

\subsection{Quantum electrodynamics}

As a  representation of the dynamics of particles and fields, quantum field theory customarily begins by specifying a Lagrangian density functional of probability amplitude fields whereas quantum computing  uses an Hamiltonian operator that generates the evolution of the state of a set of qubits. Beginning with the Lagrangian density functional for a 4-spinor Dirac field in quantum electrodynamics, the free  Lagrangian density for the Dirac field $\psi(x)$ for a quantum particle of mass $m$ is
\begin{equation}
\label{free_Dirac_particle_Lagrangian}
{\cal L}^\text{Dirac}[\psi] = \overline{\psi}(x) (i \gamma^\mu \partial_\mu-m)\psi(x),
\end{equation}
where $\overline{\psi}(x) \equiv \psi^\dagger(x) \gamma^0$, and  where   $\gamma^\mu=(\gamma_0, \bm{\gamma})$ has Dirac matrix components in the chiral representation
\begin{equation}
{\gamma}_0 = \sigma_x\otimes\bm{1}
\qquad
\qquad
\bm{\gamma} = i\sigma_y \otimes\bm{\sigma}. 
\end{equation}
The free  Lagrangian density for the massless fermion field (chiral Weyl particle)  is obtained by taking $m\rightarrow 0$. 
   Much  in quantum field theory regarding the dynamics of the Dirac field $\psi(x)$ follows from ${\cal L}^\text{Dirac}[\psi]$.  The matter-gauge field interaction in the theory may be obtained from (\ref{free_Dirac_particle_Lagrangian})  by  replacing the 4-derivative with a generalized 4-derivative that includes a spatially-dependent  4-potential field $A_\mu(x)$, with a radiation part that mediates the interaction between separated Dirac particles and a background part that otherwise steers  the Dirac particle along a curvilinear trajectory. That is, by applying the prescription for the gauge covariant derivative  $\partial_\mu \mapsto  \partial_\mu - i e A_\mu(x)/(\hbar c)$ (so that ${\cal L}^\text{Dirac}[ \psi] \mapsto {\cal L}^\text{Dirac}[ \psi, A]$), (\ref{free_Dirac_particle_Lagrangian}) becomes the Lagrangian density for a Dirac particle moving in a 4-potential field.  
  To complete the interaction dynamics,  equations of motion for the 4-potential field itself must be modeled, so  to (\ref{free_Dirac_particle_Lagrangian}) is added a covariant Lagrangian density  for the gauge field 
\begin{equation}
\label{Maxwell_Lagrangian_density}
  {\cal L}^\text{Maxwell}[A] = -\frac{1}{4} F_{\mu\nu}(x)F^{\mu\nu}(x),
\end{equation}  
   where the field strength tensor is $F^{\mu\nu}(x) = \partial^\mu A^\nu(x) - \partial^\nu A^\mu(x)$.
   Then, for example, all the vertex factors in the Feynman diagrams for the particle-particle interactions needed in perturbation expansions of the quantum field theory  follow from ${\cal L}[\psi, A] = {\cal L}^\text{Dirac}[\psi, A] + {\cal L}^\text{Maxwell}[A]$. 
   The action is $S[\psi,A] = \int d^4 x\, {\cal L}[\psi,A]$,
where  $\int d^4 x\equiv c\int dt \int d^3x$.\footnote{ 
The quantity $S[\psi,A]/\hbar$ is dimensionless and contains all the phase information of the quantum system. 
Denoting the physical units of length $L$, mass $M$, and time $T$, the dimensions of the action are $[S]= ML^2/T$. Any Lagrangian constructed from quantum fields must have dimensions of $[{\cal L}/(\hbar c)]=L^{-4}$, the physical dimensions of inverse 4-density. 
}
The Euler-Lagrange equations
\begin{equation}
\label{Euler_Lagrange_equation}
\partial_\mu\left(\frac{\partial {\cal L}[\psi,A]}{\partial(\partial_\mu \psi)}\right)
-
\frac{\partial{\cal L}[\psi,A]}{\partial \psi}
=
0,
\qquad
\partial_\mu
\left(
\frac{\partial{{\cal L}[\psi,A]}}{\partial(\partial_\mu A_\nu)} 
\right)
-
\frac{\partial{{\cal L}[\psi,A]}}{\partial A_\nu}
=0
\end{equation}
  are obtained by varying the action with respect to $\psi$ and $A^\mu$ and setting $\delta S[\psi,A]=0$.
Therefore, by adding (\ref{Maxwell_Lagrangian_density}) to model the dynamics of the Maxwell field,  ${\cal L}^\text{Dirac}[ \psi, A]$ becomes the Lagrangian density for a Dirac particle moving in a  4-potential field and also interacting with other Dirac particles via the transverse part of the 4-potential field.

\subsubsection{QED in 4-vector notation}

The QED Lagrangian density is
\begin{equation}
\label{QED_Lagrangian_density}
{\cal L}[\gamma, \psi, A] =
 i \overline{\psi}(x)\gamma^\mu\partial_{\mu}\psi(x)-e\overline{\psi}\gamma^{\mu}A_\mu(x)\psi(x) - m \overline{\psi}(x)\psi(x)
-\frac{1}{4}F_{\mu\nu}(x)F^{\mu\nu}(x),
\end{equation}
with  the  Maxwell field is $A^\mu = (A_0, \bm{A})$ and source field  is $e J^\mu = e \overline{\psi}\gamma^\mu \psi$, where $\mu,\nu = 0,1,2,3$. 
In (\ref{QED_Lagrangian_density}), the Lagrangian density functional's dependence on the  Dirac gamma matrices $\gamma^\mu$ is explicitly indicated, ${\cal L} = {\cal L}[\gamma, \psi, A] $.   In component form, the 4-vector fields appearing in the theory (\ref{QED_Lagrangian_density}) are
\begin{subequations}
\begin{eqnarray}
A^\mu & = & (A_0, A_x, A_y, A_z)=(A_0, \bm{A})
\\
J^\mu & = & (J_0, J_x, J_y, J_z)=(\rho, \bm{J}).
\end{eqnarray}
\end{subequations}
The Euler-Lagrange equations (\ref{Euler_Lagrange_equation}) respectively give the equations of motion
\begin{subequations}
\label{QED_equations_of_motion}
\begin{eqnarray}
\label{Dirac_equation_in_field}
(i \gamma^\mu\partial_{\mu} - m )\psi(x) &=& e \gamma^{\mu}A_\mu(x)\psi(x)
\\
\label{Maxwell_equations_with_source}
\partial_{\mu}F^{\mu\nu}(x) &=&  e \overline{\psi}(x)\gamma^\nu\psi(x) .
\end{eqnarray}
\end{subequations}
Denoting the probability current density field as $J^\nu\equiv \overline{\psi}(x)\gamma^\nu\psi(x)$, the  set of coupled  Dirac-Maxwell equations  
may be written as
\begin{subequations}
\label{Dirac_Maxwell_equations_of_motion}
\begin{eqnarray}
i\hbar c \,\gamma_\mu \left(\partial^\mu  - i \frac{eA^\mu}{\hbar c}\right) \psi
& =&
    mc^2\psi
\\
\label{Maxwell_equations_1}
F^{\mu\nu} &=&\partial^\mu A^\nu - \partial^\nu A^\mu
\\
\label{Maxwell_equations_2}
e J^\nu&=&\partial_\mu F^{\mu\nu}
\\
\partial_\nu J^\nu 
&=&
0.
\end{eqnarray}
\end{subequations}
The last equation for  4-current density conservation  follows from $\partial_\nu\partial_\mu F^{\mu\nu}=0$, which vanishes because the derivative ordering is symmetric  under interchange of indices whereas the field tensor is antisymmetric.

\subsubsection{Dirac-Maxwell equations in  tensor-product notation}

Using 4-spinor and tensor-product notation, the coupled Dirac-Maxwell's equations for the 4-spinor matter field $\psi$,  4-spinor potential  field ${\cal A}$,  4-spinor electromagnetic  field ${\cal F}$, and   4-spinor current density (source) field ${\cal J}$, 
\begin{equation}
\psi
=
 {\scriptsize
\begin{pmatrix}
   \psi_{\text{\tiny L} \uparrow}   \\
       \psi_{\text{\tiny L} \downarrow} \\
          \psi_{\text{\tiny R} \uparrow} \\
          \psi_{\text{\tiny R} \downarrow}
\end{pmatrix}
}
\quad
{\cal A} 
 =  
 {\scriptsize
\begin{pmatrix}
      -A_x + i A_y    \\
  A_0+   A_z \\
   -A_0+   A_z\\
     A_x + i A_y  \\
\end{pmatrix}
}
\quad
{\cal F }
 =  
 {\scriptsize
\begin{pmatrix}
     - F_x + i F_y    \\
       - \partial\cdot A  + F_z \\
        \partial\cdot A + F_z \\
      F_x + i F_y  
\end{pmatrix}
}
\quad
{\cal J }
 = 
 {\scriptsize
\begin{pmatrix}
      -J_x + i J_y    \\
   \rho  +J_z\\
    -\rho +J_z\\
     J_x + i J_y  \\
\end{pmatrix}
},
\end{equation}
are equivalently specified by
\begin{subequations}
\label{Maxwell_equation_4spinor_rep_covariant_form}
\begin{eqnarray}
\label{Dirac_equation_matrix_chiral_form}
\begin{pmatrix}
  - m   &  i \sigma \cdot (\partial -i e A) \\
  i \bar{\sigma} \cdot (\partial-ieA)    & - m
\end{pmatrix}
\begin{pmatrix}
      \psi_\text{\tiny L}    \\
       \psi_\text{\tiny R}  
\end{pmatrix}
&=&
 0
 \\
\label{Maxwell_equation_4spinor_rep_covariant_form_a}
-{\cal F} &=&
 \bm{1}\otimes   \bar\sigma\cdot \partial {\cal A}, 
\\
\label{Maxwell_equation_4spinor_rep_covariant_form_b}
-e {\cal J} &=&
 \bm{1}\otimes \sigma\cdot \partial {\cal F},
\end{eqnarray}
\end{subequations}
where
$\sigma^\mu = (1,\bm{\sigma})$, $\bar\sigma^\mu=(1,-\bm{\sigma})$,
 $\sigma\cdot \partial = \sigma^\mu \partial_\mu$, and $\bar\sigma\cdot \partial = \bar\sigma^\mu \partial_\mu$.
 The form of the Dirac equation in (\ref{Dirac_equation_matrix_chiral_form}) is conventional \cite{Peskin_Schroeder_2004}, and a derivation of the Maxwell equations in the form of   (\ref{Maxwell_equation_4spinor_rep_covariant_form_a}) and (\ref{Maxwell_equation_4spinor_rep_covariant_form_b}) 
 is given in Appendix~\ref{A_mu_field_dynamics}.

\subsection{Superconducting fermionic fluid}
\label{Sec_Superconducting_quantum_fluid}

The connection between Bose-Einstein condensation and superfluidity and superconductivity were originally discovered by London \cite{PhysRev.54.947,PhysRev.74.562}.  The dynamical behavior of a superconducting  fluid comprised of Dirac particles may be described by a QED-like gauge field theory with a massive 4-potential field.  The set of the equations of motion are similar to Maxwell's equations with sources that are electrically charged Dirac particles.  The difference is that the Maxwell field becomes a massive bosonic field with London mass $m_\text{\tiny $L$}
=
\frac{\hbar}{c}   \sqrt{\frac{ e^2\rho}{mc^{2}}}$, where   $\rho = (\psi^\dagger \psi)_\circ$ is the probability density. The equations of motion for a superconducting quantum fluid herein referred to as the Dirac-Maxwell-London equations.  These equations are presented here using several different notations, including  the conventional  4-vector notation and a novel tensor-product notation based on paired  4-spinors.  The  tensor-product notation  is subsequently used  (below in Section~\ref{Section_quantum_lattice_gas_model}) to write the quantum lattice gas algorithm for  quantum superconducting fluid dynamics, and in turn the quantum algorithm for quantum electrodynamics in the limit where the London penetration depth approaches the size of the system.

\subsubsection{Dirac-Maxwell-London  equations in 4-vector notation}

For a superconductor, the charge 4-current density is related to the 4-potential as
\begin{equation}
\label{London_J_A_identity}
\lambda_\text{\tiny $L$}^2 e J^\mu =-A^\mu,
\end{equation}
so the  Dirac-Maxwell-London  equations of motion become
\begin{subequations}
\label{Dirac_Maxwell_London_equations_standard_4_vector_form}
\begin{eqnarray}
i\hbar c \,\gamma_\mu \left(\partial^\mu  - i \frac{eA^\mu}{\hbar c}\right) \psi
 &=&
    mc^2\psi
\\
\label{Maxwell_London_equations_1}
F^{\mu\nu} &=&\partial^\mu A^\nu - \partial^\nu A^\mu
\\
\label{Maxwell_London_equations_2}
-\frac{1}{\lambda_\text{\tiny $L$}^2} A^\nu&=&\partial_\mu F^{\mu\nu}
\\
\partial_\nu A^\nu 
&=&
0.
\end{eqnarray}
\end{subequations}

\subsubsection{Dirac-Maxwell-London  equations in  tensor-product notation}

Alternatively, for a superconducting fluid we may write (\ref{London_J_A_identity}) as
\begin{equation}
e {\cal J} =-m_\text{\tiny $L$}^2 {\cal A},
\end{equation}
and the Dirac-Maxwell-London equations (\ref{Dirac_Maxwell_London_equations_standard_4_vector_form}) as
\begin{subequations}
\label{Dirac_Maxwell_London_equations}
\begin{eqnarray}
\begin{pmatrix}
  - m   &  i \sigma \cdot (\partial -i e A) \\
  i \bar{\sigma} \cdot (\partial-ieA)    & - m
\end{pmatrix}
\begin{pmatrix}
      \psi_\text{\tiny L}    \\
       \psi_\text{\tiny R}  
\end{pmatrix}
&=&
 0
\\
\label{Maxwell_London_equations_in_Yepez_form}
\begin{pmatrix}
  - m_\text{\tiny $L$}    &  i \bm{1}\otimes\sigma \\
  i \bm{1}\otimes\bar{\sigma} \cdot \partial    & - m_\text{\tiny $L$}
\end{pmatrix}
\begin{pmatrix}
      {\cal A}    \\
     \tilde {\cal A}
\end{pmatrix}
&=&
 0,
\end{eqnarray}
\end{subequations}
for  4-spinor potential fields
\begin{equation}
\label{zeroth_generation_Majorana_spin_1_potential_fields}
{\cal A}
=
{\scriptsize
\begin{pmatrix}
      -A_x + i A_y    \\
  A_0+   A_z \\
   -A_0+   A_z\\
     A_x + i A_y  \\
\end{pmatrix}
}
,
\qquad
\tilde {\cal A}
=
-\frac{i}{m_\text{\tiny $L$} }
 {\scriptsize
\begin{pmatrix}
     - F_x + i F_y    \\
       - \partial\cdot A  + F_z \\
        \partial\cdot A + F_z \\
      F_x + i F_y  
\end{pmatrix}
}.
\end{equation}
Maxwell's equations for the  4-potential field are recovered from (\ref{Maxwell_London_equations_in_Yepez_form}) in the $m_\text{\tiny $L$}\rightarrow 0$ limit.\footnote{
One must take care in taking the $m_\text{\tiny $L$}\rightarrow 0$ limit of (\ref{Maxwell_London_equations_in_Yepez_form}).  The first component equation of (\ref{Maxwell_London_equations_in_Yepez_form}) should be multiplied by $m_\text{\tiny $L$}$ before taking the limit to avoid a trivial singularity. Furthermore, it is not necessary to set the London mass equal to zero.  It is sufficient to set the London rest energy equal to the highest energy scale, $m_\text{\tiny $L$} c^2= \hbar/\tau$ limit.} 
This 4-spinor representation of the Maxwell-London equations (\ref{Maxwell_London_equations_in_Yepez_form}) is derived in Appendix~\ref{Sec_derivation_Maxwell_London_equations_Yepez_form}.

\subsubsection{Dirac-Maxwell-London  equations in paired 4-spinor notation}

Introducing an 8-component local state $\Phi$ as a pair of 4-spinor fields
\begin{equation}
\Phi = 
\begin{pmatrix}
      \Phi_1    \\
       \Phi_2  
\end{pmatrix}
=
 \frac{\Phi_\circ\ell}{e}
\begin{pmatrix}
      {\cal A}    \\
      \tilde {  \cal A}
\end{pmatrix},
\end{equation}
(\ref{Dirac_Maxwell_London_equations}) may be written in a manifestly covariant way using a coupled pair of Dirac equations 
\begin{subequations}
\label{Dirac_Maxwell_London_equations_symmetrical_4_vector_form}
\begin{eqnarray}
i\hbar c \,\gamma_\mu \left(\partial^\mu  - i \frac{eA^\mu}{\hbar c}\right) \psi
-
    mc^2\psi
 &=&
 0
    \\
  \left( \frac{m_\text{\tiny $L$} c}{\hbar}\right)^2 A_\mu &=& -e \overline{\psi}(x)\gamma_\mu\psi(x)
  \\
\label{Maxwell_London_equations_symmetrical_4_vector_form}
i\hbar c \,{\cal G}_\mu \partial^\mu  \Phi
-
    m_\text{\tiny $L$}c^2\Phi
     &=&
     0,
\end{eqnarray}
\end{subequations}
where   ${\cal G}^\mu=({\cal G}_0, \bm{{\cal G}})$ has generalized Dirac matrix components, and where in the chiral representation
\begin{equation}
{\cal G}_0 = \sigma_x\otimes \bm{1} \otimes\bm{1}
\qquad
\bm{{\cal G}} = i\sigma_y \otimes\bm{1} \otimes\bm{\sigma}. 
\end{equation}
The paired 4-spinor generalized Dirac equation representation of the Dirac-Maxwell-London equations (\ref{Maxwell_London_equations_symmetrical_4_vector_form}) are derived in Appendix~\ref{Sec_derivation_Maxwell_London_equations_Yepez_generalized_Dirac_form}, and it is a  concise way to express all the gauge field dynamics in a single equation.
The  Lagrangian density used here for a superconducting quantum fluid of fermions with a novel four-point interaction is
\begin{equation}
\label{fermionic_superconducting_Lagrangian_denstity}
\begin{split}
{\cal L}[\psi, A,  \Phi]
&
 =
 i \hbar c \overline{\psi}(x)\gamma_\mu\left(\partial^{\mu}+\frac{i e}{\hbar c} A^\mu(x)\right)\psi(x) - m c^2\overline{\psi}(x)\psi(x)
 - \frac{1}{2} 
 \lambda_\text{\tiny $L$}^2\,
\overline{\psi}(x) [\gamma^\mu, \gamma^\nu] \psi(x) |\Phi_\circ|^2 \ell^2 A_\mu(x)  A_\nu(x)
 \\
 & +
 i \hbar c  \overline{\Phi}(x){\cal G_\mu}\partial^{\mu}\Phi(x) -  m_\text{\tiny $L$} c^2\overline{\Phi}(x)\Phi(x) .
 \end{split}
\end{equation}
The Dirac-Maxwell-London equations  (\ref{Dirac_Maxwell_London_equations_symmetrical_4_vector_form}) are obtained by varying with respect to $\psi$,  $A^\mu$, and $\Phi$ (in the second line).

 The Lagrangian density functional for 
a Landau-Ginzburg bosonic field $\phi$ interacting with a 
Maxwell field $A^\mu$ is
\begin{subequations}
\label{Higgs_model}
\begin{eqnarray}
{\cal L}_\text{Landau-Ginzburg}[\phi,A] & = & {\cal L}_\text{interacting-massless K.G.}[\phi, A] 
+
{\cal L}_\text{Maxwell}[A] +{\cal L}_\text{nonlinear}[\phi]
\label{phi_Higgs_Lagrangian_density}
\\
 & = & \hbar c \left|{\cal D}_\mu \phi\right|^2
  -\left. \frac{1}{4} \middle( \partial^\mu A^\nu -   \partial^\nu A^\mu\right)^2
  - V(\phi),
\end{eqnarray}
where there is minimal coupling between the $\phi$ and $A^\mu$ fields via
${\cal D}_\mu  = \partial_\mu + i \frac{e}{\hbar c } A_\mu(x)$, 
and there is a nonlinear self-coupling for the $\phi$ field via
${\cal L}_\text{nonlinear}[\phi]
=
-V(\phi)
 = \mu^2 |\phi|^2 - \frac{\lambda}{2}|\phi|^2$. 
\end{subequations}
Theory (\ref{Higgs_model}) represents a type-II superconductor (i.e. a superconductor with magnetic quantum vortices).  The Maxwell field $A_\mu$ acquires a mass  (say $m_\text{\tiny L}$ via the Higgs mechanism), so the $A_\mu$ field can penetrate into a superconductor only up to a depth of $m_\text{\tiny L}^{-1}$ (the well known Meissner effect).  In the simple case (Abelian gauge group) when $\phi$ is a complex scalar field, then the $U(1)$ gauge symmetry is
\begin{equation}
\phi(x) \rightarrow e^{i \alpha(x)} \phi(x),
\qquad
A_\mu(x) \rightarrow A_\mu(x) - \frac{\hbar c}{e}\partial_\mu \alpha(x).
\end{equation}
The  Landau-Ginzberg Lagrangian density (\ref{Higgs_model}), for the bosonic Cooper pair field, is also known as the Abelian Higgs model.   In 1+1 dimensions, the  bosonic sine-Gordon model is the dual of the fermionic Thirring model \cite{PhysRevD.11.2088,thirring_58}.    
In 3+1 dimensions, the Abelian  Langrangian density  (\ref{Higgs_model}) may  be considered  a dual bosonic theory of the  fermionic Lagrangian density (\ref{fermionic_superconducting_Lagrangian_denstity}).

\section{Quantum lattice gas method}
\label{Section_quantum_lattice_gas_model}

\subsection{$\psi$-$A^\mu$ interaction}

In a quantum lattice gas model, how  one treats  source terms on the righthand side in (\ref{QED_equations_of_motion}) 
 depends on the interpretation of
 the  interaction Lagrangian, which in QED with minimal coupling is
\begin{equation}
\label{basic_interaction_Lagrangian}
{\cal L}^\text{\tiny int}[\psi,A] =-e\overline{\psi}(x)\gamma_{\mu}A^\mu(x)\psi(x).
\end{equation}
 The electron-photon interaction in QED occurs at a vertex point, which is commonly depicted by the Feynman diagram
\begin{equation}
\label{Feynman_vertex_reaction}
\includegraphics[width=1in]{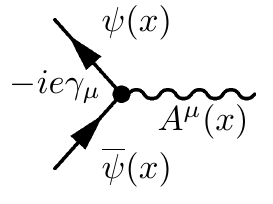}.
\end{equation}
In a quantum lattice gas,  (\ref{Feynman_vertex_reaction}) is interpreted as a reversible reaction, modeled by a unitary collide operator acting on the qubits
  representing the Dirac field $\psi(x)$
and the qubits representing the 4-potential field $A^\mu(x)$. 
There are two distinct interaction mechanisms. 
The first is a forward reaction  whereby the $A^\mu(x)$ field changes the $\psi(x)$ field. The second is a back reaction whereby the $\psi(x)$ field changes the $A^\mu(x)$ field.  

Using dimensional units for $\hbar$ and $c$, the forward reaction  represented by (\ref{basic_interaction_Lagrangian}) can be expressed as a unitary transformation of the $\psi(x)$ field
\begin{equation}
\label{psi_A_forward_reaction}
\psi'(x)
=
e^{-i\ell  \gamma_\mu \frac{e A^\mu(x)}{\hbar c}}\psi(x) .
\end{equation}
The back reaction can be calculated as a transformation of the probability current $J^\nu(x) = \overline{\psi}(x)\gamma^\nu \psi(x)$---the source on the righthand side of (\ref{Maxwell_equations_with_source})---that can  be expressed as a unitary transformation of the 4-potential field $A^\nu(x)$. 
To derive this unitary transformation, one can begin by expanding (\ref{psi_A_forward_reaction}) to first order
\begin{equation}
\label{psi_A_forward_reaction_expansion}
\psi'(x)
=
\psi(x) 
-
i\ell  \gamma_\mu \frac{e A^\mu(x)}{\hbar c} \psi(x)
+
\cdots.
\end{equation}
Then, upon making use of the adjoint gamma matrices $\gamma^{\mu\dagger}$ and the anticommutation relation $\{\gamma^{\mu\dagger}, \gamma^{\nu\dagger}\}=2\eta^{\mu\nu}$,  an expansion for the outgoing probability current density $J'^\nu(x)=\overline{\psi'}(x)  \gamma_\nu \psi'(x)$ is
\begin{subequations}
\label{Outgoing_current_density_expansion}
\begin{eqnarray}
J'^\nu(x)
&=&
\psi'^\dagger(x)\gamma_0  \gamma_\mu \psi'(x)
\\
&\stackrel{(\ref{psi_A_forward_reaction_expansion})}{=}&
\left(
\psi^\dagger(x) 
+
i\ell  \frac{e A^\mu(x)}{\hbar c} \psi^\dagger(x) \gamma_\mu^\dagger 
+
\cdots
\right)
\gamma_0  \gamma^\nu
\left(
\psi(x) 
-
i\ell  \gamma_\kappa \frac{e A^\kappa(x)}{\hbar c} \psi(x)
+
\cdots
\right)
\\
&=&
\psi^\dagger(x)\gamma_0  \gamma^\nu \psi(x)
+
i\ell  \frac{e A^\mu(x)}{\hbar c} \psi^\dagger(x)
 \gamma_\mu^\dagger 
\gamma_0  \gamma^\nu
\psi(x) 
-
i\ell 
  \frac{e A^\kappa(x)}{\hbar c} 
\psi^\dagger(x) 
\gamma_0  \gamma^\nu
 \gamma_\kappa
  \psi(x)
+
\cdots
\\
&=&
J^\nu(x)
+
\left.
i\ell 
  \frac{e A^\mu(x)}{\hbar c} 
\psi^\dagger(x) 
\middle(
 \gamma_\mu^\dagger
\gamma_0  
\gamma^\nu
-
\gamma_0  
\gamma^\nu
 \gamma_\mu
\right)
  \psi(x)
+
\cdots
\\
&=&
J^\nu(x)
+
\left.
i\ell 
  \frac{e A_\mu(x)}{\hbar c} 
\psi^\dagger(x) 
\middle(
2\eta^{0\mu}
\gamma^\nu
-
\gamma_0 
 \gamma^{\mu\dagger}
 \gamma^\nu
-
\gamma_0  
\gamma^\nu
 \gamma^\mu 
\right)
  \psi(x)
+
\cdots
\\
\label{Outgoing_current_density_expansion_f}
&=&
J^\nu(x)
+
\left.
i\ell 
  \frac{e A_\mu(x)}{\hbar c} 
\psi^\dagger(x) 
\middle(
2\eta^{0\mu}
\gamma^\nu
-
\gamma_0
\left(
 \gamma^{\mu\dagger} 
 \gamma^\nu
+
  \gamma^\mu \gamma^\nu
  -
  [ \gamma^{\mu},  \gamma^\nu]
 \right)
 \right)
  \psi(x)
+
\cdots
\\
\label{Outgoing_current_density_expansion_g}
&=&
J^\nu(x)
+
\left.
i\ell 
  \frac{e A_\mu(x)}{\hbar c} 
\psi^\dagger(x) 
\middle(
\left(
2\eta^{0\mu}
-
 \gamma_0
 \gamma^{\mu\dagger} 
 -
  \gamma_0
  \gamma^\mu 
\right)
\gamma^\nu
-
\gamma_0
  [ \gamma^{\mu},  \gamma^\nu]
  \right)
  \psi(x)
+
\cdots
\\
\label{Outgoing_current_density_expansion_h}
&=&
J^\nu(x)
-
i\ell 
  \frac{e A_\mu(x)}{\hbar c} 
\psi^\dagger(x) \gamma_0
  [ \gamma^{\mu},  \gamma^\nu]
  \psi(x)
+
\cdots.
\end{eqnarray}
\end{subequations}
Furthermore, with the  identity $A^\mu(x) = - e \lambda_\text{\tiny L}^2 J^\nu(x) = -\frac{mc^2}{e\rho}J^\nu(x)$, the analytical expansion of the back reaction is obtained
\begin{subequations}
\label{psi_A_backward_reaction_expansion}
\begin{eqnarray}
\label{psi_A_backward_reaction_expansion_a}
A'^\nu(x)
&=&
A^\nu(x)
+
i 
  \frac{  mc^2\ell }{\hbar c \rho} 
{\psi}^\dagger(x)
 \gamma_0
[ \gamma^\mu ,
 \gamma^\nu
 ]
  \psi(x)
A_\mu(x)
+
\cdots
\end{eqnarray}
\end{subequations}
Here the outgoing  local value of the Maxwell field $A'^\mu(x)$ is determined by the local values of the incoming Dirac field $\psi^\dagger(x)$, the outgoing Dirac field $\psi(x)$, and the incoming value of the Maxwell field $A^\mu(x)$.

\subsection{Discrete spacetime representation}

Since the back reaction part of the interaction involves the emission or absorption of a quanta of  radiation  (e.g. a photon in QED),  only the transverse 4-potential components need be considered.  Therefore, one may take $A_0(x)=0$ in the back reaction formula when it is applied.  
In any case, with the time derivative approximated as $\partial_t\psi \approxeq (\psi' - \psi)/\tau$ and with speed of light $c=\ell/\tau$, the Dirac-Maxwell-London equations (\ref{Dirac_Maxwell_London_equations_symmetrical_4_vector_form})  written in update form are
\begin{subequations}
\label{Dirac_Maxwell_London_equations_symmetrical_update_form}
\begin{eqnarray}
\psi'(x) 
 &=&
\psi(x)  + \ell \gamma_0 \bm{\gamma} \cdot  \left(\nabla
 + 
 i \frac{e\bm{A}(x)}{\hbar c}\right) \psi(x) 
- i  \frac{mc^2\tau}{\hbar} \gamma_0 \psi(x) 
+
i  \frac{e A_0(x) \tau}{\hbar} \psi(x) 
+
\cdots
    \\
\label{Dirac_Maxwell_London_equations_symmetrical_update_form_b}
A'^\mu(x) 
&\stackrel{(\ref{psi_A_backward_reaction_expansion_a})}{=}&
A^\mu(x)
-
i 
  \frac{2  mc^2\tau }{\hbar} 
\frac{\overline{\psi}(x) 
[ \gamma^\mu ,\gamma^\nu]
  \psi(x)}{\rho}
+
\cdots
\\
\Phi'(x) 
 &=&
\Phi(x)  + \ell {\cal G}_0\cdot \bm{{\cal G}} \cdot  \nabla \Phi(x) 
- i  \frac{m_\text{\tiny $L$}c^2\tau}{\hbar} {\cal G}_0 \Phi(x)  
+
\cdots,
\end{eqnarray}
\end{subequations}
If we take the limit as  ${\hbar}/{(m_\text{\tiny $L$} c)}$ approaches the  size $L$ of the system (or $m_\text{\tiny $L$}\rightarrow 0$), then (\ref{Dirac_Maxwell_London_equations_symmetrical_update_form}) becomes the update rule for the equations of motion (\ref{Dirac_Maxwell_equations_of_motion}) of quantum electrodynamics.  In this limit, the QED fields $\psi(x)$ and $A^\mu(x)$ may be modeled as constituent fields of a superconducting quantum fluid with London penetration depth equal to the size of the system (i.e. the size of the universe). 

The quantum lattice gas model is a unitary representation of the equations of motion of gauge field theory.  So  the expansions (\ref{Dirac_Maxwell_London_equations_symmetrical_update_form}) are taken to be  low-energy expansions of the dynamical equations of motion expressed in manifestly unitary form on a spacetime lattice with cell sizes $\ell$ and $\tau$
\begin{subequations}
\label{Dirac_Maxwell_London_equations_symmetrical_unitary_form}
\begin{eqnarray}
\label{Dirac_Maxwell_London_equations_symmetrical_unitary_form_psi_evolution}
\psi(x+\ell \gamma)
 &=&
e^{ \ell \gamma_0 \bm{\gamma} \cdot  \left(\nabla
 + i \frac{e \bm{A}(x)}{\hbar c}\right) 
- i  \frac{mc^2\tau}{\hbar} \gamma_0 
+
i  \frac{e A_0(x) \tau}{\hbar}
} \psi(x)
    \\
\label{Dirac_Maxwell_London_equations_symmetrical_unitary_form_psi_A_collision}
A'^\mu(x) 
&=&
e^{
   - i  
  \frac{ 2 mc^2\tau }{\hbar} 
\frac{\overline{\psi}(x) 
[ \gamma^\mu ,\gamma^\nu]
  \psi(x)}{\rho}
}
A_\nu(x) 
\\
\label{Dirac_Maxwell_London_equations_symmetrical_unitary_form_Phi_evolution}
\Phi(x+\ell {\cal G})
 &=&
e^{\ell  {\cal G}_0 \bm{{\cal G}} \cdot  \nabla
- i  \frac{m_\text{\tiny $L$}c^2\tau}{\hbar} {\cal G}_0 
}\Phi(x) .
\end{eqnarray}
\end{subequations}
The evolution equation (\ref{Dirac_Maxwell_London_equations_symmetrical_unitary_form_psi_evolution})  describes the dynamical behavior of the Dirac particle (solid fermion line in the Feynman diagram (\ref{Feynman_vertex_reaction})), keeping the 4-potential field fixed. The evolution equation (\ref{Dirac_Maxwell_London_equations_symmetrical_unitary_form_psi_A_collision}) includes the local emission (absorption) of a quanta of radiation 
 at a vertex point, with the particles otherwise not moving. The evolution equation (\ref{Dirac_Maxwell_London_equations_symmetrical_unitary_form_Phi_evolution})  describes the dynamical behavior of the 4-potential field (wavy gauge field line in the Feynman diagram (\ref{Feynman_vertex_reaction})), keeping the matter field fixed.  

The set of evolution equations (\ref{Dirac_Maxwell_London_equations_symmetrical_unitary_form}) serve as the basis for the quantum lattice gas  split-operator representation of gauge field theory---an Abelian quantum field theory in this example.  A quantum lattice gas algorithm based on (\ref{Dirac_Maxwell_London_equations_symmetrical_unitary_form}) is presented in the next section.

\subsection{Stream and collide operators}
\label{Stream_collide_update_rule}

In writing (\ref{Dirac_Maxwell_London_equations_symmetrical_unitary_form}),  $\psi$, $A$, and $\Phi$ were taken to be continuous and differentiable probability amplitude fields defined on a continuous spacetime manifold.   Yet, a qubit array can exactly represent (\ref{Dirac_Maxwell_London_equations_symmetrical_unitary_form}), and thereby it can also represent a differentiable field defined on a continuous and differentiable spacetime manifold. This remarkable property of the qubit array follows from the  momentum operator  mapped to a spacetime derivative,  $ \hat p^\mu \mapsto i \hbar \partial^\mu$.  Motion on the qubit array is represented by $e^{i \ell \gamma_\mu \hat p^\mu/\hbar}$ as a unitary operator that shifts a field value stored in the qubits at point $x_\mu$ to a field value stored in the neighboring qubits at $x_\mu+ \ell \gamma_\mu$.\footnote{The qubit array encodes both the spacetime and the particles contained therein. So one may add the matrix-valued  quantity $\ell \gamma_\mu$ to  $x_\mu$  because the position ket   $|x, q_1,\cdots, q_Q\rangle$  at a point contains  all the state information in a $2^Q$ dimensional local Hilbert space.  This encoding is explained in Sec.~\ref{Section_Many_fermions_on_a_qubit_array} below.
}
Motion in the continuous spacetime picture is represented by a stream operator ${\cal S} = e^{\ell \gamma_\mu \partial^\mu}$,  an unitary operation that shifts a field value at a point $x_\mu$ to  the nearby point $x_\mu+ \ell \gamma_\mu$.  Switching between lattice  and continuous space pictures is a commonly employed practice here, akin to the use of continuous quantum fields in solid-state physics for describing many-body particle dynamics in crystallographic lattices.  The continuous wave picture for particle motion in a crystal may be referred to as the Bloch-Wannier picture, and this  is outlined in Appendix~\ref{Sec_Bloch_Wannier_continuous_field_picture}.

The quantum state at a point  is denoted  by $\psi(x)$ and $\Phi(x)$  (that contains ${\cal A}(x)$) and the values of these fields constitute the local state at $x$.  
  The local time-dependent evolution equation of motion is expressed as a rule that simultaneously updates the local state of each  point in the system.    In the simplest model, the update rule may be written as a  product of a stream step ${\cal S}$ and a collide step ${\cal C}$ 
\begin{subequations}
\label{Simplest_QLG_model_SC_map}
\begin{eqnarray}
\label{Simplest_QLG_model_SC_map_psi}
 \psi'(x) 
 &=&
  {\cal S}( \gamma, A) {\cal C}(\gamma,m) \psi(x) \mapsto \psi(x)
  \\
  {\cal A}'(x) 
 &=&
{\cal C}[\psi] {\cal A}(x) \mapsto {\cal A}(x)   
\\
 \Phi'(x) 
 &=&
\label{Simplest_QLG_model_SC_map_Phi}
  {\cal S}( {\cal G}, 0) {\cal C}({\cal G},m_\text{\tiny $L$}) \Phi(x) \mapsto \Phi(x),
\end{eqnarray}
\end{subequations}
where the   collide operator ${\cal C}[\psi]= 
   \text{exp}\!\left(
   - i  
   2 mc^2
   U
\theta[\psi]
U^{\text{\tiny c}\dagger}
\tau/\hbar 
 \right)
   $
is   the nonlinear unitary transformation 
 appearing in (\ref{Dirac_Maxwell_London_equations_symmetrical_unitary_form_psi_A_collision}), where $\theta[\psi]$ is the $4\times 4$ matrix representation of ${\overline{\psi}(x) 
[ \gamma^\mu ,\gamma^\nu]
  \psi(x)}/{\rho}
$, and where the unitary matrices $U$ and $U^\text{\tiny c}$ transform from 4-vectors to 4-spinor forms ${\cal A} =  U A^\mu$ and ${\cal A} = U^\text{\tiny c} A_\mu$.
 ${\cal S}(\gamma, A)$  shifts the components from $\psi(x)$ to  $\psi(x')$ for $x'^\mu= x^\mu + \ell \gamma^\mu$  in the neighborhood of $x^\mu$ while it  performs a unitary gauge group transformation $e^{i \ell \gamma_0 \cdot \gamma^\mu  e A_\mu(x)/(\hbar c)}$. 
  The stream-collide update rule (\ref{Simplest_QLG_model_SC_map_psi}), or likewise (\ref{Simplest_QLG_model_SC_map_Phi}), is written with the understanding that the collide operator 
is applied simultaneously to all points of the system, and then 
 the stream operator 
 is applied simultaneously to all points of the system, thereby completing one iteration of the evolution for the respective field.
  In the quantum field theory limit, the collide operator is represented by the unitary matrix \cite{yepez_arXiv1512.02550_quant_ph}
\begin{subequations}
\label{Dirac_equation_collide_operator_QFT_limit}
\begin{eqnarray}
{\cal C}(\gamma,m)
&\approxeq  &
  \begin{pmatrix}
\sqrt{1-\left(\frac{mc^2\tau}{\hbar}\right)^2}\,\bm{1}
  &
-i\frac{mc^2\tau}{\hbar} \,\bm{1}
  \cr
-i\frac{mc^2\tau}{\hbar}\,\bm{1}& 
\sqrt{1-\left(\frac{mc^2\tau}{\hbar}\right)^2}\,\bm{1}
\end{pmatrix}
\\
{\cal C}({\cal G},m_\text{\tiny $L$})
&\approxeq  &
  \begin{pmatrix}
\sqrt{1-\left(\frac{m_\text{\tiny $L$}c^2\tau}{\hbar}\right)^2}\,\bm{1}\otimes\bm{1}
  &
-i\frac{m_\text{\tiny $L$}c^2\tau}{\hbar} \,\bm{1}\otimes\bm{1}
  \cr
-i\frac{m_\text{\tiny $L$}c^2\tau}{\hbar}\,\bm{1}\otimes\bm{1}& 
\sqrt{1-\left(\frac{m_\text{\tiny $L$}c^2\tau}{\hbar}\right)^2}\,\bm{1}\otimes\bm{1}
\end{pmatrix}.
\end{eqnarray}
\end{subequations}
The update rule (\ref{Simplest_QLG_model_SC_map})  may be written as  local equations of motion
\begin{subequations}
\label{Simplest_QLG_model}
\begin{eqnarray}
\label{Simplest_QLG_model_Sdagger_C_form_psi}
{\cal S}^\dagger( \gamma, A) \psi(x) 
& =&  
{\cal C}(\gamma,m) \psi(x)
\\
\label{Simplest_QLG_model_Sdagger_C_form_Phi}
{\cal S}^\dagger( {\cal G}, 0) \Phi(x) 
& =&  
{\cal C}({\cal G},m_\text{\tiny $L$}){\cal C}[\psi] \Phi(x).
\end{eqnarray}
\end{subequations}
Furthermore, (\ref{Simplest_QLG_model_Sdagger_C_form_psi}) and (\ref{Simplest_QLG_model_Sdagger_C_form_Phi})  may be written by replacing the adjoint stream operators ${\cal S}^\dagger$ by the spacetime lattice displacements they represent
\begin{subequations}
\label{Simplest_QLG_model_shift_C_form}
\begin{eqnarray}
\psi(x-\ell \gamma - i  \ell \gamma_o  \gamma \cdot  eA(x)/(\hbar c)) 
&=&
 {\cal C}(\gamma,m) \psi(x)
 \\
\Phi(x-\ell {\cal G}) 
&=&
{\cal C}({\cal G},m_\text{\tiny $L$})  {\cal C}[\psi]\Phi(x).
\end{eqnarray}
\end{subequations}
The quantum lattice gas model (\ref{Simplest_QLG_model})  describes fermion dynamics on a spacetime lattice that is congruent to the particle dynamics governed by quantum wave equation (\ref{Dirac_Maxwell_London_equations_standard_4_vector_form}) (or equivalently (\ref{Dirac_Maxwell_London_equations_symmetrical_4_vector_form}))
 in Minkowski space.
  To help explain the mathematical basis of this congruency, which primarily derives from the application of the streaming operator, as an example one can examine the simplest case when $m=0$ and $A^\mu=0$; the basic stream operator applied to $\psi$ for this example is explained in Appendix~\ref{Streaming_protocol}.

\section{Quantum lattice gas algorithm}
\label{QEQ_quantum_algorithm}

\subsection{Quantum algorithm for the Dirac-Maxwell-London and QED equations}

  In the quantum lattice gas algorithm for chiral particle motion in a 4-potential field,  we approximate the stream operators as 
\begin{subequations}
\begin{eqnarray}
 {\cal S}(\gamma, A) 
 &\approxeq  &
 e^{i\frac{e A_0(x) \tau}{\hbar}} 
 {\cal S}_ x(\gamma, A) {\cal S}_y(\gamma, A) {\cal S}_z(\gamma, A)
 \\
  {\cal S}({\cal G}, 0) 
 &\approxeq  &
 {\cal S}_ x({\cal G}, 0) {\cal S}_y({\cal G}, 0) {\cal S}_z({\cal G}, 0),
\end{eqnarray}
\end{subequations}
 where $\gamma^0= \sigma_x\otimes\bm{1}$ and $\bm{\gamma}=i\sigma_y\otimes \sigma_z$, 
  where ${\cal G}^0= \sigma_x\otimes\bm{1}\otimes\bm{1}$ and $\bm{{\cal G}}=i\sigma_y \otimes\bm{1}\otimes\sigma_z$, 
  and 
where the stream operators along the $i$th direction are
\begin{subequations}
\begin{eqnarray}
{\cal S}_i(\gamma, A)
&=&
e^{ \ell  \sigma_z\otimes
 \sigma_i 
 \partial_i}
e^{ i \ell  \sigma_z\otimes
 {\sigma_i} \frac{e \bm{A}_i(x)}{\hbar c}
}
\\
{\cal S}_i({\cal G},0)
&=&
e^{ \ell  \sigma_z\otimes\bm{1}\otimes
 \sigma_i 
 \partial_i}
,
\end{eqnarray}
\end{subequations}
 for $i=x,y,z$.  Breaking the chiral symmetry, in the quantum lattice gas algorithm for a massive Dirac particle moving in a 4-potential field, the collide operators (\ref{Dirac_equation_collide_operator_QFT_limit}) may be written as
\begin{subequations}
\label{collision_operators_for_psi_and_Phi_in_the_QFT_limit}
\begin{eqnarray}
 {\cal C}(\gamma,\epsilon) 
 &=&
  \sqrt{1-\epsilon^2}\,\bm{1}_4
-
i\epsilon
\sigma_x\otimes\bm{1}
\\
 {\cal C}({\cal G}, \epsilon_\text{\tiny $L$}) 
 &=&
  \sqrt{1-\epsilon_\text{\tiny $L$}^2}\,\bm{1}_8
-
i\epsilon
\sigma_x\otimes\bm{1}_4
\end{eqnarray}
\end{subequations}
 in the QFT limit \cite{yepez_arXiv1512.02550_quant_ph} for $\epsilon = mc^2\tau/\hbar$ and $\epsilon_\text{\tiny $L$} = m_\text{\tiny $L$}c^2\tau/\hbar$.
  Hence,  the quantum lattice gas  evolution operator for the $\psi(x)$ field and including the forward reaction is
expressed by the  unitary algorithmic protocol
\begin{equation}
\label{Uqlg_Dirac_propagator}
{\cal U}_\psi= e^{i\frac{e A_0(x) \ell}{\hbar c}} 
{\cal S}_{x}(\gamma, A){\cal S}_{y}(\gamma, A){\cal S}_{z}(\gamma, A,m){\cal C}(\gamma,\epsilon)
\approxeq
e^{-i \frac{h_\text{D}(\gamma, A,m) \ell}{\hbar c}},
\end{equation}
where 
\begin{subequations}
\label{Streaming_algorithm_based_on_classical_shifts}
\begin{eqnarray}
{\cal S}_{x}(\gamma, A)
&=&
e^{-i\frac{\pi}{4}\bm{1}\otimes\sigma_y} 
\cdot 
e^{ \ell  \sigma_z\otimes
\sigma_z
 \partial_x}
 \cdot
 e^{ i \ell  \sigma_z\otimes
 {\sigma_z} \frac{e \bm{A}_x(x)}{\hbar c}
}
  \cdot
   e^{i\frac{\pi}{4}  \bm{1}\otimes\sigma_y}
   \qquad
   \\
{\cal S}_{y}(\gamma, A)
  & =&
   e^{i\frac{\pi}{4}  \bm{1}\otimes\sigma_x}
    \cdot
e^{ \ell  \sigma_z\otimes
\sigma_z
 \partial_y}
 \cdot
 e^{ i \ell  \sigma_z\otimes
 {\sigma_z} \frac{e \bm{A}_y(x)}{\hbar c}
}
      \cdot
       e^{-i\frac{\pi}{4} \bm{1} \otimes  \sigma_x} 
       \\
{\cal S}_{z}(\gamma, A)
  & =&
e^{ \ell  \sigma_z\otimes
\sigma_z
 \partial_z}
 \cdot
 e^{ i \ell  \sigma_z\otimes
 {\sigma_z} \frac{e \bm{A}_z(x)}{\hbar c}
}   ,
 \end{eqnarray}
\end{subequations}
 implemented  on the cubic lattice in terms of  diagonal stream operators $e^{ \ell  \sigma_z\otimes
\sigma_z
 \partial_i}$.  The hermitian generator of the evolution of $\psi$ is the Dirac Hamiltonian 
\begin{equation}
\label{Derived_Dirac_Hamiltonian_in_QFT_limit}
{{h}}_D(\gamma, A,m)
=
-\sigma_z \otimes\bm{\sigma}\cdot \left(-i\hbar c\nabla- e \bm{A}(x)\right)
+
\sigma_x\otimes\bm{1} \, mc^2 - eA_0.
\end{equation}
The quantum algorithm (\ref{Uqlg_Dirac_propagator}) is the same as the previous quantum algorithm for a relativistic Dirac 4-spinor field  \cite{yepez-qip-05}, where according to (\ref{Streaming_algorithm_based_on_classical_shifts}) all partical streaming can be implemented with classical shifts $e^{\ell \sigma_z\sigma_z \nabla}$  but where  now   a  gauge-field induced phase rotation  $e^{\ell \sigma_z\sigma_z \frac{e \bm{A}(x)}{\hbar c}}$ is included.  

  Furthermore,  the quantum lattice gas  evolution operator for the $\Phi(x)$ field and including the back reaction is
expressed by the unitary algorithmic protocol
\begin{equation}
\label{Uqlg_Dirac_propagator_Phi}
{\cal U}_\Phi= 
{\cal S}_{x}({\cal G}, 0){\cal S}_{y}({\cal G}, 0){\cal S}_{z}({\cal G}, 0)
 {\cal C}({\cal G}, \epsilon_\text{\tiny $L$})  {\cal C}[\psi]
\approxeq
e^{-i \frac{h_\text{D}({\cal G}, \psi ,m_\text{\tiny $L$}) \ell}{\hbar c}} ,
\end{equation}
where 
\begin{subequations}
\begin{eqnarray}
{\cal S}_{x}({\cal G}, 0)
&=&
e^{-i\frac{\pi}{4}\bm{1}\otimes\bm{1}\otimes\sigma_y} 
\cdot 
e^{ \ell  \sigma_z\otimes
\bm{1}\otimes\sigma_z
 \partial_x}
  \cdot
   e^{i\frac{\pi}{4}  \bm{1}\otimes\bm{1}\otimes\sigma_y}
   \qquad
   \\
{\cal S}_{y}({\cal G}, 0)
  & =&
   e^{i\frac{\pi}{4}  \bm{1}\otimes\bm{1}\otimes\sigma_x}
    \cdot
e^{ \ell  \sigma_z\otimes
\bm{1}\otimes
\sigma_z
 \partial_y}
      \cdot
       e^{-i\frac{\pi}{4} \bm{1}\otimes \bm{1}\otimes  \sigma_x} 
       \\
{\cal S}_{z}({\cal G}, 0)
  & =&
e^{ \ell  \sigma_z\otimes
\bm{1}\otimes
\sigma_z
 \partial_z}
,
 \end{eqnarray}
\end{subequations}
 implemented  on the cubic lattice in terms of  diagonal operators $e^{ \ell  \sigma_z \otimes\bm{1}\otimes
\sigma_z
 \partial_i}$.   The hermitian generator of the evolution of $\Phi$ is a generalized Dirac Hamiltonian 
\begin{equation}
\label{Derived_Dirac_Hamiltonian_in_QFT_limit}
{{h}}_D({\cal G}, \psi,m_\text{\tiny $L$})
=
-\sigma_z\otimes \bm{1} \otimes\bm{\sigma}\cdot \left(-i\hbar c\nabla\right)
+
\sigma_x\otimes\bm{1}\otimes \bm{1}\, m_\text{\tiny $L$} c^2
+
   2 mc^2  U  \theta[\psi]  U^{\text{\tiny c}\dagger}
   ,
\end{equation}
 where $\theta[\psi]$ is the $4\times 4$ matrix representation of ${\overline{\psi}(x) 
[ \gamma^\mu ,\gamma^\nu]
  \psi(x)}/{\rho}
$. 
 The  state of the system at time $t$ may be expressed as a tensor product $\Psi(t)\equiv \bigotimes_{\bm{x}\in \text{grid}} \psi(x) \otimes \Phi(x)$. 
   Accounting for the nonlocal physics associated with quantum entanglement, the simple update rule (\ref{Simplest_QLG_model}) is  fully expressed as a unitary evolution equation for the system ket
 \begin{equation}
\label{QLG_dynamical_equation_of_motion_S_C_form}
\Psi(t+\tau)
\stackrel{(\ref{Uqlg_Dirac_propagator_Phi})}{\stackrel{(\ref{Uqlg_Dirac_propagator})}{=}} 
 \Big[\bigotimes_{\bm{x}\in \text{grid}}{\cal U}_\Phi \otimes {\cal U}_\psi \Big] \Psi(t).
\end{equation}
Now, if  long-range quantum entanglement exists in the system, then one can raise the question of whether or not it is appropriate to evolve $\Psi(t)$ with the tensor-product unitary operator $\bigotimes_{\bm{x}\in \text{grid}} {\cal U}_\Phi  {\cal U}_\psi$ 
written in (\ref{QLG_dynamical_equation_of_motion_S_C_form}). 
The answer is that this  tensor-product form of unitary evolution is complete in the sense that 
 (\ref{QLG_dynamical_equation_of_motion_S_C_form}) can represent all the particle dynamics otherwise described by gauge field theories, including 
 all nonlocal particle physics. 
   Equation (\ref{QLG_dynamical_equation_of_motion_S_C_form})  is an example quantum lattice gas equation of motion that models an Abelian gauge field theory on a qubit array.

\subsection{Lorentz invariant quantum lattice gas model}

The quantum lattice models for the Maxwell-London equations for a superconducting fluid and for quantum electrodynamics  rely on unitary evolution operators (\ref{Uqlg_Dirac_propagator}) and (\ref{Uqlg_Dirac_propagator_Phi}) that are generated by a Dirac Hamiltonian and generalization thereof, respectively.  If one considers the quantum behavior of the modeled Dirac fields at the highest energy scales (approaching the grid scale of the spacetime lattice), then there appears a time dilation effect due to the form of the collision operator that derives from the path summation representation of quantum particle dynamics on the spacetime lattice \cite{yepez_arXiv1512.02550_quant_ph} 
\begin{subequations}
\label{collision_operators_for_psi_and_Phi}
\begin{eqnarray}
 {\cal C}^\text{\tiny H.E.}(\gamma,\epsilon) 
 &=&
  \sqrt{1-\epsilon^2}\,\bm{1}_4
-
i\epsilon
\sigma_x\otimes\bm{1}
 \cdot
e^{ i\ell
\sigma_z \otimes \bm{\sigma}\cdot 
\nabla
}
=
 {\cal C}(\gamma,\epsilon) +\cdots
\\
 {\cal C}^\text{\tiny H.E.}({\cal G},\epsilon_\text{\tiny $L$}) 
 &=&
  \sqrt{1-\epsilon_\text{\tiny $L$}^2}\,\bm{1}_8
-
i\epsilon
\sigma_x\otimes\bm{1}\otimes\bm{1}
 \cdot
e^{ i\ell
\sigma_z \otimes\bm{1} \otimes\bm{\sigma}\cdot 
\nabla
}
=
 {\cal C}({\cal G}, \epsilon_\text{\tiny $L$}) 
 +\cdots.
 \end{eqnarray}
\end{subequations}
This reduces to (\ref{collision_operators_for_psi_and_Phi_in_the_QFT_limit}) in the low-energy limit (when the Compton wavelength of the Dirac particle is much larger than the grid scale).  Using (\ref{collision_operators_for_psi_and_Phi}) in the quantum algorithm for the evolution of the $\psi$ and $\Phi$ fields are
\begin{subequations}
\label{quantum_lattice_algoarithm_for_psi_and_Phi_highest_energy}
\begin{eqnarray}
\psi(\bm{x}, t+ \tau)
&=&
{\cal S}_{x}(\gamma, A){\cal S}_{y}(\gamma, A){\cal S}_{z}(\gamma, A)
 {\cal C}^\text{\tiny H.E.}(\gamma,\epsilon)
\psi(\bm{x},t)
\\
\Phi(\bm{x}, t+ \tau)
&=&
{\cal S}_{x}({\cal G}, 0){\cal S}_{y}({\cal G}, 0){\cal S}_{z}({\cal G}, 0)
 {\cal C}^\text{\tiny H.E.}({\cal G},\epsilon_\text{\tiny $L$})  {\cal C}[\psi]  \Phi(\bm{x},t)
.
\end{eqnarray}
\end{subequations}
These can be written in manifestly unitary form \cite{yepez_arXiv1512.02550_quant_ph} 
\begin{subequations}
\label{quantum_lattice_algoarithm_for_psi_and_Phi_highest_energy_unitary}
\begin{eqnarray}
\psi(\bm{x}, t+ \tau)
&=&
e^{-i \frac{h_\text{D}(\gamma, A,m) \zeta \tau}{\hbar}}
\psi(\bm{x},t)
\\
\Phi(\bm{x}, t+ \tau)
&=&
e^{-i  \frac{h_\text{D}({\cal G}, \psi,m_\text{\tiny $L$})\zeta_\text{\tiny $L$} \tau}{\hbar}}
\Phi(\bm{x},t),
\end{eqnarray}
\end{subequations}
where the  unitary evolution includes   dimensionless scale factors $1\le \zeta \le \frac{\pi}{2}$ and $1\le  \zeta_\text{\tiny $L$} \le \frac{\pi}{2}$.  The appearance of $\zeta$ (and likewise for $\zeta_\text{\tiny $L$}$) implies that  the smallest observable intervals are the  distance $r=\zeta\ell$ and elapsed time $t_r =  r/c = \zeta \tau$.   Yet, with particular relevance to modeling gauge field theories, the appearance of the scale factor implies exact Lorentz invariance in the quantum lattice model at all scales.  This is a welcomed feature of the lattice model.   Making the modeling task simpler, in applications such as model low-energy gauge field theories such as QED, the scale factor becomes unity.  So for QED, the quantum lattice gas algorithm takes the simpler form
\begin{subequations}
\label{quantum_lattice_algoarithm_for_psi_and_Phi_QED}
\begin{eqnarray}
\psi(\bm{x}, t+ \tau)
&=&
{\cal S}_{x}(\gamma, A){\cal S}_{y}(\gamma, A){\cal S}_{z}(\gamma, A)
 {\cal C}(\gamma,\epsilon) \psi(\bm{x},t)
\\
\Phi(\bm{x}, t+ \tau)
&=&
{\cal S}_{x}({\cal G}, 0){\cal S}_{y}({\cal G}, 0){\cal S}_{z}({\cal G}, 0)
 {\cal C}[\psi]   \Phi(\bm{x},t)  
.
\end{eqnarray}
\end{subequations}
The algorithmic steps needed to run the quantum lattice gas algorithm are: (1)  initialize the $\psi$ and $\Phi$ fields; (2) update the $\psi$ field; (3) update the $\Phi$ field; (4) time advances by one time step $\tau$; and (5) to continue to evolve in time, go to Step 2.
%

\subsection{Many-fermion quantum simulation on a qubit array}
\label{Section_Many_fermions_on_a_qubit_array}

With $Q$ qubits per point in the qubit array, the ket $|\Psi\rangle$ at a point $x$ is a $2^Q$-multiplet ket
\begin{eqnarray}
|\Psi\rangle 
&=&
 \sum_{q_1=0}^{1}  \sum_{q_2=0}^{1} \cdots  \sum_{q_Q=0}^{1} \Psi_{q_1, q_2, \cdots, q_Q} |q_1, q_2, \cdots, q_Q\rangle
=\sum_{N=0}^{2^Q-1} \Psi_N |N\rangle,
\end{eqnarray}
where in the second line the binary encoded index is
$N= 2^{Q-1} q_1 + 2^{Q-2} q_{2} +\cdots  + 2 q_{Q-1} + q_Q$, 
for Boolean number variables $q_\alpha = 0, 1$ for $\alpha=1, 2, \cdots, Q$. 
With $V=L^3$ number of points on a grid of size $L$, the total number of qubits in the qubit array is $V Q$.  
Each point in space is assigned a position-basis ket, denoted by $|x, N\rangle$, in a large but finite Hilbert space of size $2^{VQ}$.
The position ket $|{x}, N\rangle=|{x}, q_1,\cdots, q_Q\rangle$ is the numbered state with all the other numbered variables at points $\ne x$  set to zero
\begin{subequations}
\label{singly_occupied_position_spin_kets_for_2_qubits_per_point}
\begin{eqnarray}
|{x}, q_1,\cdots, q_Q\rangle & = & | 0000\quad  \cdots \underbrace{{q_1, q_2, \cdots, q_Q}}_{\text{point $x$}} \cdots \quad 0000 \rangle.
\end{eqnarray}
\end{subequations}
Therefore, in this position representation, a probability amplitude (a complex number) at a point is $\Psi_{q_1, \cdots, q_Q}(x) = \langle {x}, q_1,\cdots, q_Q | \Psi\rangle$, which can be written concisely as $\Psi_N(x) = \langle {x},N | \Psi\rangle$. 
The position kets are orthonormal
$\langle  {x}_a, N | {x}_b, N' \rangle = \delta_{ab} \delta_{N N'}$. 
The notion of a point in position space and the local quantum state at that point are intrinsically linked in the quantum lattice gas method.

  Acting on a system of $Q$ qubits, $  a^\dagger_\alpha$ and $  a_\alpha$ create and destroy a fermionic number variable at the $\alpha\hbox{th}$ qubit
\begin{eqnarray}
\nonumber
  a^\dagger_\alpha |q_1  \dots q_\alpha \dots q_Q\rangle & = & 
\left\{
\begin{matrix}
0&, &  q_\alpha=1\cr
\varepsilon\; |q_1 \dots 1 \dots q_Q\rangle&, &   q_\alpha=0\cr
\end{matrix}
\right. ,
\quad
  a_\alpha |q_1  \dots q_\alpha \dots q_Q\rangle  = 
\left\{
\begin{matrix}
\varepsilon\; |q_1  \dots 0 \dots q_Q\rangle&, & q_\alpha=1\cr
0&, & q_\alpha=0\cr
\end{matrix}
\right. ,
\\
\end{eqnarray}
where the phase factor is \cite{fetter-71}
$\varepsilon = (-1)^{\sum_{i=1}^{\alpha-1}n_i}$. 
The fermionic ladder operators satisfy  anticommutation relations
\begin{eqnarray}
\label{fermionic_anti_commutation_relations}
\{   a_\alpha,   a^\dagger_\beta\} & =&  \delta_{\alpha\beta}
\qquad
\{   a_\alpha,   a_\beta\}  =   0
\qquad
\{   a^\dagger_\alpha,   a^\dagger_\beta\}  =  0.
\end{eqnarray}
The number operator $  n_\alpha \equiv   a^\dagger_\alpha   a_\alpha$ has eigenvalues of 1 or 0  in the number representation when acting on a pure state, corresponding to the $\alpha\hbox{th}$ qubit being in state $|1\rangle$ or $|0\rangle$ respectively.  With the logical one 
\(
{\scriptsize
|1\rangle = \begin{pmatrix}
     0    \\
      1  
\end{pmatrix}
}
\)  
and logical zero
\(
{\scriptsize
|0\rangle = \begin{pmatrix}
     1    \\
      0  
\end{pmatrix}
} 
\) 
states of a qubit,
notice that 
\(
\sigma_z |1\rangle = -|1\rangle,
\)
so one can count the number of preceding bits that contribute to the overall phase shift due to fermionic bit exchange involving the $i$th qubit with tensor product operator, 
 \(
 \sigma_z^{\otimes i-1} |\Psi \rangle = (-1)^{N_i}| \Psi\rangle.
\)
The phase factor is determined by the number of bit crossings 
\(
N_i = \sum_{k=1}^{i-1}q_k
\)
in the state $|\Psi\rangle$
and where the Boolean number variables are $q_k\in [ 0,1]$.
Hence, an annihilation operator is decomposed into a tensor product known as the Jordan-Wigner transformation \cite{JordanWigner1928}
$a_\alpha =\sigma_z^{\otimes \alpha-1} \otimes\, a\otimes \bm{1}^{\otimes Q-\alpha}$, 
for integer $\alpha \in [1,Q]$.

 With $Q=4$, the local ket 
 $|\Psi\rangle$ at a point $x$  can encode two Dirac 4-spinors $\psi(x)$ and $\tilde\psi(x)$ as well as two  4-spinor fields ${\cal A}(x)$ and $\tilde {\cal A}(x)$
\begin{subequations}
\begin{equation}
{\cal A}
=
\begin{pmatrix}
   {\cal A}_{\text{\tiny L} \uparrow}   \\
       {\cal A}_{\text{\tiny L} \downarrow} \\
          {\cal A}_{\text{\tiny R} \uparrow} \\
          {\cal A}_{\text{\tiny R} \downarrow}
\end{pmatrix}
\qquad
\tilde{{\cal A}}
=
\begin{pmatrix}
  \tilde{{\cal A}}_{\text{\tiny L} \uparrow}   \\
       \tilde{{\cal A}}_{\text{\tiny L} \downarrow} \\
          \tilde{{\cal A}}_{\text{\tiny R} \uparrow} \\
         \tilde{{\cal A}}_{\text{\tiny R} \downarrow}
\end{pmatrix}
\quad
\psi
=
\begin{pmatrix}
   \psi_{\text{\tiny L} \uparrow}   \\
       \psi_{\text{\tiny L} \downarrow} \\
          \psi_{\text{\tiny R} \uparrow} \\
          \psi_{\text{\tiny R} \downarrow}
\end{pmatrix}
\qquad
\tilde{\psi}
=
\begin{pmatrix}
  \tilde{\psi}_{\text{\tiny L} \uparrow}   \\
       \tilde{\psi}_{\text{\tiny L} \downarrow} \\
          \tilde{\psi}_{\text{\tiny R} \uparrow} \\
         \tilde{\psi}_{\text{\tiny R} \downarrow}
\end{pmatrix}
\end{equation}
 using the four qubits, say  $|q_g q_l q_o q_s\rangle$, as 
\begin{equation}
\label{4_Dirac_4_spinor_fields_Q4_encoding}
\begin{split}
 |q_g q_l q_o q_s\rangle
 &= 
 {\cal A}_{\text{\tiny L} \uparrow}|0000\rangle+{\cal A}_{\text{\tiny L} \downarrow}|0001\rangle+ {\cal A}_{\text{\tiny R} \uparrow}|0010\rangle+{\cal A}_{\text{\tiny R} \downarrow}|0011\rangle
 +
 \tilde{{\cal A}}_{\text{\tiny R} \downarrow}|0111\rangle+\overline{{\cal A}}_{\text{\tiny R} \uparrow}|0110\rangle+ \tilde{{\cal A}}_{\text{\tiny L} \downarrow}|0101\rangle+\tilde{{\cal A}}_{\text{\tiny L} \uparrow}|0100\rangle
\\
 &+ 
 \psi_{\text{\tiny L} \uparrow}|1000\rangle+\psi_{\text{\tiny L} \downarrow}|1001\rangle+ \psi_{\text{\tiny R} \uparrow}|1010\rangle+\psi_{\text{\tiny R} \downarrow}|1011\rangle
 +
 \tilde{\psi}_{\text{\tiny R} \downarrow}|1111\rangle+\tilde{\psi}_{\text{\tiny R} \uparrow}|1110\rangle+ \tilde{\psi}_{\text{\tiny L} \downarrow}|1101\rangle+\tilde{\psi}_{\text{\tiny L} \uparrow}|1100\rangle.
 \end{split}
\end{equation} 
\end{subequations}

The  qubit $|q_g\rangle$ encodes the particle's generation, where $|q_g\rangle = |0\rangle$ for the zeroth generation (4-spinors ${\cal A}$ and $\tilde{{\cal A}}$) and $|q_g\rangle = |1\rangle$ for the first generation (Dirac 4-spinors $\psi$ and $\tilde{\psi}$).  The unitary operators (\ref{Uqlg_Dirac_propagator}) and (\ref{Uqlg_Dirac_propagator_Phi}) can be implemented using quantum gate operations  represented in terms of the fermionic qubit annihilation and creation operators $ a_\alpha$ and $a^\dagger_\alpha= a_\alpha^\text{T}$ acting on the qubit array.  This approach serves as  a general quantum computational formulation applicable to any quantum lattice gas algorithm for a many-fermion system.

\subsection{Path integral representation of quantum field theory}

The probability amplitude for a particle to go from point $a$ to point $b$ on the qubit array is calculated using a kernel operator
\begin{equation}
\label{yepez_feynman_path_summation_spin_form4}
\hat{K}^\text{\tiny HE}_{ab}
=
\sum_{\bm{n}}
\frac{1}{(\ell L)^3}
e^{i\frac{{\bm{x}}\cdot {\bm{p}_{\bm{n}}}}{\hbar}}
\sum_{\{{\bm{s}}_1, \dots ,{\bm{s}}_{N-1}\}}
\!\!\!\!\! \ell^3
\;
e^{-\frac{i}{\hbar}  \sum_{w=0}^{N-1} \delta t \,{\hat{h}}_\text{\tiny QLG} },
\end{equation}
where set of spin chains $\{{\bm{{s}}}_1,\dots,{\bm{{s}}}_{N-1}\}$   enumerate paths $\ell\sum_{w=0}^{N-1} {\bm{s}}_w
= \bm{x}_b-\bm{x}_a$ 
with fixed endpoints $\bm{x}_a$ and $\bm{x}_b$, and where
 $
\sum_{\bm{n}}\equiv \sum_{n_{x} = -L/2}^{(L/2)-1} \sum_{n_{y} = -L/2}^{(L/2)-1} \sum_{n_{z} = -L/2}^{(L/2)-1}$ sums over all  grid momenta  ${\bm{p}}_{{\bm{n}}} \equiv
 \frac{2\pi \hbar}{L}\left(\frac{n_x}{{\ell}}, \frac{n_y}{{\ell}}, \frac{n_z}{{\ell}}\right)$  for a grid of size $L$ \cite{yepez_arXiv1512.02550_quant_ph}. 
In (\ref{yepez_feynman_path_summation_spin_form4}), the time differential is $\delta t= \zeta \tau$ for scale factor $\zeta$ and the quantum lattice gas Hamiltonian operator is
\begin{equation}
\label{QLG_Hamiltonian_operator}
 {\hat{h}}_\text{\tiny QLG} 
 =
-  \Gamma^0
\bm{\Gamma} \cdot {\bm{\hat{p}}}_{\bm{n}} c  -   \Gamma^0 Mc^2 +  X -  \Gamma^0\bm{\Gamma} \cdot \bm{Y}.
\end{equation}
 With singleton qubit number and hole operators $n$ and $h$, 
the generalized Dirac  matrices $\Gamma^\mu = (\Gamma_0, \bm{\Gamma})$ are
\begin{equation}
\Gamma^0
\equiv 
n\otimes\bm{1}\otimes\gamma^0
+
h\otimes{\cal G}^0,
\qquad
\bm{\Gamma}
\equiv
 n\otimes\bm{1}\otimes \bm{\gamma}
+
 h\otimes \bm{{\cal G}}.
\end{equation}
In (\ref{QLG_Hamiltonian_operator}), the diagonal mass operator is $M = n\otimes \bm{1}_8\, m + h\otimes \bm{1}_8 \, m_\text{\tiny L}$, and   $X$ and $\bm{Y}$ are nonlinear interaction generators 
\begin{equation}
 X =  
n\otimes h \otimes \bm{1}_4 \, e A_0
+   h\otimes h\otimes 2 mc^2  U  \theta[\psi]  U^{\text{\tiny c}\dagger},
\qquad
\bm{Y} = n\otimes h \otimes \bm{1}_4 \, e \bm{A}.
\end{equation}

 In Minkowski space, one can convert to the Bloch-Wannier picture by taking $\ell \sim dx\sim dy\sim dz$ and $\frac{2\pi}{L\ell}\sim \frac{dp}{\hbar}$, 
so the summations over $\bm{n}=(n_x, n_y, n_z)$ and spin configurations $\{{\bm{s}}_1, \dots ,{\bm{s}}_{N-1}\}$ in 
(\ref{yepez_feynman_path_summation_spin_form4}) map  to momentum-space and path integrals, respectively,
\begin{eqnarray}
\nonumber
\sum_{\bm{n}}\frac{1}{(2\pi)^3} \left(\frac{2\pi}{L\ell}\right)^3 
\sum_{\{{\bm{s}}_1, \dots ,{\bm{s}}_{N-1}\}}
  \!\!\! \!\!\!\ell^3
&\sim&
\int  \frac{d^3p}{(2\pi \hbar)^3} \int_a^b {\cal D}\{ \bm{x}(t) \}.
\label{path_integral_map}
\end{eqnarray}
With  stream operator ${\hat{\cal S}}(\bm{n}) 
=e^{ i\ell \sigma_z
 \bm{\sigma}\cdot \hat{\bm{p}}_{\bm{n}} /\hbar
}$
expressed with the momentum operator $\hat{\bm{p}}_{\bm{n}}  = \frac{\hbar \bm{n}\cdot\bm{\sigma}}{L\ell}$,  one converts to the Bloch-Wannier picture by replacing the  grid operators $\hat{\bm{p}}_{\bm{n}}$ and $\hat{E}_{\bm{n}} =\hat{\bm{p}}_{\bm{n}} c^2/v$  on the qubit array with derivative operators acting on a  Dirac spinor field $\psi(x)$ in continuous spacetime, i.e.
$\bm{\hat{p}}_{\bm{n}} \sim -i\hbar\nabla$ 
 and 
$\hat{E}_{\bm{n}} \sim i\hbar\partial_t$. 
The long wavelength limit is $\ell {k}_{\bm{n}} \lll  1$ and the low rest energy limit is  $mc^2\tau/\hbar \lll 1$. 
Writing $\hat{E}_{\bm{n}}=\hat{\bm{p}}_{\bm{n}} c^2/v = \dot{\bm{\hat{x}}}\cdot\bm{\hat{p}}_{\bm{n}}+\bm{\hat{x}}\cdot\dot{\bm{\hat{p}}}_{\bm{n}}$ 
in the 
 Bloch-Wannier picture, the kernel (\ref{yepez_feynman_path_summation_spin_form4})     is congruent to a path integral
\begin{subequations}
\label{feynman_path_integral_in_p_space}
\begin{eqnarray}
\hat{K}^\text{\tiny HE}_{ab}
&\stackrel{(\ref{path_integral_map})}{\cong}&
\int\frac{dp^3}{(2\pi\hbar)^3}
\int_a^b {\cal D}\{ \bm{x}(t)\} \,
e^{i\frac{{\bm{x}}\cdot {\bm{p}}}{\hbar}}
 e^{-\frac{i}{\hbar} \int dt\, {\hat{h}}_\text{\tiny QLG} }  
=
\int_a^b {\cal D}\{ \bm{x}(t)\} 
\int\frac{dp^3}{(2\pi\hbar)^3}
\,
e^{\frac{i}{\hbar} \int dt\, {\hat{L}_\text{\tiny QLG} }},
\end{eqnarray}
\end{subequations}
where  the Lagrangian operator ${\hat{L}_\text{\tiny QLG} }$ is determined  by 
\begin{equation}
\label{Lagrangian_operator_Legendre_form}
{\hat{L}_\text{\tiny QLG} }
\equiv
\hat{E}_{\bm{n}}-  {\hat{h}}_\text{\tiny QLG} 
 =
 \Gamma^0
 \left[(\Gamma^0 
\hat{E}_{\bm{n}}- c\bm{\Gamma} \cdot {\bm{\hat{p}}}_{\bm{n}})  -   Mc^2 +  \Gamma^0 X - \bm{\Gamma} \cdot \bm{Y} \right].
\end{equation}
Finally, the Lagrangian density is expressed in terms of the Lagrangian operator as ${\cal L}_\text{\tiny QLG} 
\equiv
\Psi^\dagger {\hat{L}_\text{\tiny QLG} } 
\Psi$, so the matrix element of (\ref{yepez_feynman_path_summation_spin_form4})  takes the form of a Feynman path integral
\begin{equation}
\label{relativistic_path_integral}
\langle \hat{K}_{ab}
\rangle
\mapsto 
\int_a^b{\cal D}\{ \bm{x}(t)\} \,
\int\frac{dp^3}{(2\pi\hbar)^3}
e^{\frac{i}{\hbar} \int d^4x \,{\cal L}_\text{\tiny QLG} }.
\end{equation}
The Lagrangian density corresponding to (\ref{Lagrangian_operator_Legendre_form}) is equivalent to (\ref{fermionic_superconducting_Lagrangian_denstity})
\begin{equation}
\label{fermionic_superconducting_Lagrangian_denstity_recovered}
\begin{split}
{\cal L}_\text{\tiny QLG} 
&
 =
 i \hbar c \overline{\psi}(x)\gamma_\mu\left(\partial^{\mu}+\frac{i e}{\hbar c} A^\mu(x)\right)\psi(x) - m c^2\overline{\psi}(x)\psi(x)
 - \frac{1}{2} 
 \lambda_\text{\tiny $L$}^2\,
\overline{\psi}(x) [\gamma^\mu, \gamma^\nu] \psi(x) |\Phi_\circ|^2 \ell^2 A_\mu(x)  A_\nu(x)
 \\
 & +
 i \hbar c  \overline{\Phi}(x){\cal G_\mu}\partial^{\mu}\Phi(x) -  m_\text{\tiny $L$} c^2\overline{\Phi}(x)\Phi(x) .
 \end{split}
\end{equation}
This completes the quantum computing picture of the many-fermion dynamics with nonlinear gauge field interactions.

\section{Conclusion}
\label{Section_Conclusion}

Quantum gas lattice  models are intended as an algorithmic scheme for programming a  lattice-based quantum computer using the quantum gate model of quantum computation \cite{feynman-85,divincenzo-pra95,divincenzo-pra95a,barenco-prsl95}.  Feynman set in place the cornerstone of quantum information
theory: a scalable quantum computer can serve as a universal quantum simulator \cite{feynman-ces60,feynman-82}. 
This Feynman conjecture that a scalable quantum computer
can simulate any other quantum system exactly is critically important because it only requires computational physical resources that scale
linearly with the spacetime volume---it does not scale exponentially with the number of particles in the modeled physical
system.  So the realization of a Feynman quantum computer would be a great breakthrough for many-body quantum physics modeling---if there are efficient quantum algorithms for modeling many-fermion quantum physics. Presented was one such quantum algorithm based on the quantum lattice gas method. Presented was the application for modeling a superconducting fluid and in turn this application led to another one presented for modeling quantum electrodynamics.    So  quantum lattice gas models can have an impact on  the foundations of quantum field theory.

In the quantum lattice model representation of many-fermion quantum systems, local unitary evolution is decomposed into  collide and stream operators
\cite{PhysRevE.63.046702}.   
The collide and stream operator method was first tested for the case  of two fermions dynamics in the nonrelativistic limit
\cite{bib:BB06}. A quantum stream operator is needed to model the relativistic quantum particle dynamics 
\cite{yepez-qip-05}.  The quantum stream and collide operators can be cast in a tensor-product representation, which is a tensor network. 
That these operators can serve as a representation of Temperley-Lieb algebra and the braid group has been established \cite{yepez:73420R}.  
The quantum stream and collide operators are implemented on a qubit array by using many-fermion annihilation and creation operators.  The many-fermion annihilation operator (and the many-fermion creation operator which is its transpose) allows for the calculation of quantum knot invariants, which are a generalization of the Jones polynomials 
\cite{PhysRevA.81.022328}. Because of previous validation tests of the quantum lattice gas method, it is expected to provide a pathway to avoid the Fermi sign problem.

Regarding future outlooks,  the quantum lattice gas algorithm can be used to investigate nonequilibrium physics in Fermi condensates of charged spin-1/2 particles.  So modeling the time-dependent behavior of strongly-correlated Fermi gases is another application.  Gauge field theories with non-Abelian gauge groups and with order unity coupling constants are not amenable to predictions by perturbative methods.  So  methods that can go beyond perturbation theory are needed.    Numerical methods such as quantum Monte Carlo methods are best suited to predicting time-independent field configurations.   Holographic techniques based on AdS/CFT correspondence between strongly coupled non-Abelian gauge field theories and weakly coupled gravitational theories hold some promise, yet such techniques have not yet proved useful for predicting time-dependent field configurations.  The possibility of predicting time-dependent field configurations  by using a quantum lattice gas algorithm for gauge field theories with non-Abelian gauge groups will be discussed in a subsequent communication.

\appendix    

\section{Derivation 4-spinor quantum equation for Maxwell  field $A^\mu(x)$}
\label{A_mu_field_dynamics}

In natural units with $c=1$, let us derive a quantum wave equation for the dynamical behavior of the Maxwell field $A^\mu_\text{\tiny T}=(0,\bm{A})$.\footnote{It is this quantum wave equation that can be readily converted into a quantum algorithm for ${\cal A}$.}  To begin, let us express the curl operator 
\begin{equation}
\nabla \times \bm{A}
=
\nabla \times 
{\scriptsize
\begin{pmatrix}
      A_x    \\
      A_y \\
      A_z  
\end{pmatrix}
}
 = 
{\scriptsize
 \begin{pmatrix}
     \partial_y A_z - \partial_z A_y
         \\
           \partial_z A_x - \partial_x A_z
 \\
           \partial_x A_y - \partial_y A_x  
\end{pmatrix}
}
=
{\scriptsize
\begin{pmatrix}
      B_x    \\
      B_y \\
      B_z  
\end{pmatrix}
}
\end{equation}
as a unitary transformation acting on  a Majorana 4-spinor field. 
The curl of the spin=1 Majorana 4-spinor  ${\cal A}$ (representing the 3-vector $\bm{A}$) should go as
\begin{equation}
\label{curl_operator_map}
{\cal A}
=
{\scriptsize 
\begin{pmatrix}
      -A_x + i A_y    \\
      A_z\\
      A_z\\
      A_x + i A_y  
\end{pmatrix}
}
 \xrightarrow{\nabla\times}
{\scriptsize 
\begin{pmatrix}
     - \partial_y A_z + \partial_z A_y + i ( \partial_z A_x - \partial_x A_z )    \\
       \partial_x A_y - \partial_y A_x\\
       \partial_x A_y - \partial_y A_x\\
      \partial_y A_z - \partial_z A_y + i (\partial_z A_x - \partial_x A_z)
\end{pmatrix}
}.
\end{equation}

\subsection{Block-diagonal representation}

One can represent the curl of the Majorana 4-spinor by the  hermitian operator
\begin{subequations}
\label{curl_operator_on_Majorana_4_spinor_block_diagonal}
\begin{eqnarray}
\nonumber
{\scriptsize 
\begin{pmatrix}
   \partial_z   &  \partial_x - i \partial_y   & 0 & 0\\
     \partial_x + i \partial_y  &  -\partial_z & 0 & 0\\
     0 & 0 & \partial_z   &  \partial_x - i \partial_y \\
       0 & 0 &   \partial_x + i \partial_y  &  -\partial_z 
\end{pmatrix}
}
{\scriptsize 
\begin{pmatrix}
      -A_x + i A_y    \\
      A_z\\
      A_z\\
      A_x + i A_y  
\end{pmatrix}
}
\\
\label{curl_operator_on_Majorana_4_spinor_block_diagonal_a}
=
{\scriptsize 
\begin{pmatrix}
      -\partial_z A_x + \partial_x A_z - i ( \partial_y A_z - \partial_z A_y )    \\
     -\nabla\cdot \bm{A} + i (  \partial_x A_y - \partial_y A_x)\\
     \nabla\cdot \bm{A} + i (  \partial_x A_y - \partial_y A_x)\\
      -\partial_z A_x + \partial_x A_z + i ( \partial_y A_z - \partial_z A_y )    
\end{pmatrix}
}
\\
\label{curl_operator_on_Majorana_4_spinor_block_diagonal_b}
= 
i
{\scriptsize 
\begin{pmatrix}
     - \partial_y A_z + \partial_z A_y + i ( \partial_z A_x - \partial_x A_z )    \\
       \partial_x A_y - \partial_y A_x + i \nabla\cdot \bm{A} \\
       \partial_x A_y - \partial_y A_x - i \nabla\cdot \bm{A} \\
      \partial_y A_z - \partial_z A_y + i (\partial_z A_x - \partial_x A_z)
\end{pmatrix}
}
\\
\label{curl_operator_on_Majorana_4_spinor_block_diagonal_c}
= 
i
{\scriptsize 
\begin{pmatrix}
     - B_x + i B_y    \\
       B_z + i \nabla\cdot \bm{A} \\
       B_z - i \nabla\cdot \bm{A} \\
      B_x + i B_y  
\end{pmatrix}
},
\end{eqnarray}
\end{subequations}
where in the last line the components of the magnetic field $\bm{B}=\nabla\times\bm{A}$ were inserted.  
The outcome (\ref{curl_operator_on_Majorana_4_spinor_block_diagonal_b}) is nearly identical to the desired result (or guess) on the righthand side of (\ref{curl_operator_map}).  
 That the hermitian curl operator also generates the divergence $\nabla\cdot \bm{A}$ is an unexpected feature of the representation.  This naturally enforces the Coulomb gauge condition 
 $\nabla\cdot \bm{A}=0$.  Furthermore, that the hermitian curl operator also multiples  the resulting magnetic field by $i$  is another unexpected but desirable feature of the representation.   This is consistent with the definition of a complex electromagnetic field  $\bm{F}= \bm{E} + i \bm{B}$.  
Then, to  produce the electric field, one simply adds a negative time derivative to each diagonal component of the hermitian operator and also includes the time component $A_0$ in the  4-spinor
\begin{subequations}
\label{4_curl_operator_on_Majorana_4_spinor_block_diagonal}
\begin{eqnarray}
\nonumber
{\scriptsize 
\begin{pmatrix}
  -\partial_t +  \partial_z   &  \partial_x - i \partial_y   & 0 & 0\\
     \partial_x + i \partial_y  &  -\partial_t   -\partial_z & 0 & 0\\
     0 & 0 &  -\partial_t +\partial_z   &   \partial_x - i \partial_y \\
       0 & 0 &   \partial_x + i \partial_y  & -\partial_t    -\partial_z 
\end{pmatrix}
}
{\scriptsize 
\begin{pmatrix}
      -A_x + i A_y    \\
      A_z + A_0\\
      A_z - A_0 \\
      A_x + i A_y  
\end{pmatrix}
}
\\
= 
{\scriptsize 
\begin{pmatrix}
      -E_x + i E_y    \\
       E_z - \partial_t A_0 \\
       E_z  + \partial_t A_0 \\
      E_x + i E_y  
\end{pmatrix}
}
+
i
{\scriptsize 
\begin{pmatrix}
     - B_x + i B_y    \\
       B_z + i \nabla\cdot {\bm A} \\
       B_z - i \nabla\cdot {\bm A}  \\
      B_x + i B_y  
\end{pmatrix}
}
\qquad
\\
= 
{\scriptsize 
\begin{pmatrix}
     - F_x + i F_y    \\
       F_z - \partial\cdot A  \\
       F_z +\partial\cdot A  \\
      F_x + i F_y  
\end{pmatrix}
},
\qquad
\end{eqnarray}
\end{subequations}
where  the electric field is $\bm{E}=-\partial_t\bm{A}-\partial_t A_0$,  and where  the complex electromagnetic field   $\bm{F}= \bm{E} + i \bm{B}$ is used in the last line. 
The equations of motion can be written as
\begin{subequations}
\label{QLG_algorithm_for_the_Maxwell_field_block_diagonal}
\begin{eqnarray}
\left(-\partial_t + \bm{\sigma}\cdot \nabla\right)
{\scriptsize
\begin{pmatrix}
      -A_x + i A_y    \\
      A_z+A_0\\
\end{pmatrix}
}
&=&
{\scriptsize 
\begin{pmatrix}
     - F_x + i F_y    \\
       F_z - \partial\cdot A   \\
\end{pmatrix}
}
\\
\left(-\partial_t + \bm{\sigma}\cdot \nabla\right)
{\scriptsize
\begin{pmatrix}
      A_z-A_0\\
      A_x + i A_y  
\end{pmatrix}
}
&=&
{\scriptsize 
\begin{pmatrix}
       F_z + \partial\cdot A \\
      F_x + i F_y  
\end{pmatrix}
}.
\end{eqnarray}
\end{subequations}
In tensor product form, the quantum wave equation for the Maxwell field (\ref{QLG_algorithm_for_the_Maxwell_field_block_diagonal}) when written in  4-spinor form is
 \begin{equation}
\label{QLG_algorithm_for_the_Maxwell_field_block_diagonal_four_spinor_form}
\left(-\partial_t +\bm{1} \otimes\bm{\sigma}\cdot \nabla\right)
{\scriptsize
\begin{pmatrix}
      -A_x + i A_y    \\
      A_0+A_z\\
      -A_0+A_z\\
      A_x + i A_y  
\end{pmatrix}
}
=
{\scriptsize 
\begin{pmatrix}
     - F_x + i F_y    \\
       - \partial\cdot A  + F_z \\
        \partial\cdot A + F_z \\
      F_x + i F_y  
\end{pmatrix}
}.
\end{equation}
  Finally, if we choose to interleave the evolution about unity, then we need to apply  the reversed-momentum operator $-\partial_t -\bm{1}\otimes \bm{\sigma}\cdot \nabla$ to both sides of (\ref{QLG_algorithm_for_the_Maxwell_field_block_diagonal_four_spinor_form}) 
 \begin{equation}
\label{Maxwell_equations_spinor_form}
\left(-\partial_t -\bm{1} \otimes\bm{\sigma}\cdot \nabla\right)
{\scriptsize
\begin{pmatrix}
     - F_x + i F_y    \\
       - \partial\cdot A  + F_z \\
        \partial\cdot A + F_z \\
      F_x + i F_y  
\end{pmatrix}
}
=
{\scriptsize 
\begin{pmatrix}
       \partial_t F_x + i (\nabla\times \bm{F})_x  - i \partial_t F_y  +(\nabla\times \bm{F})_y + (\partial_x - i \partial_y)\partial\cdot A\\
    - \partial_t F_z -   i (\nabla\times \bm{F})_z  +  \nabla\cdot \bm{F}  +   (\partial_t -  \partial_z)\partial\cdot A\\
  - \partial_t F_z -   i (\nabla\times \bm{F})_z  -  \nabla\cdot \bm{F} -   (\partial_t +  \partial_z)\partial\cdot A\\
  -\partial_t F_x - i (\nabla\times \bm{F})_x  - i \partial_t F_y  + (\nabla\times \bm{F})_y  - (\partial_x + i \partial_y)\partial\cdot A\\
\end{pmatrix}
}
\equiv
 e {\cal J},
\end{equation}
where (with $\partial\cdot A=0$) the righthand side defines the source 4-spinor field (associated with the 4-current $eJ^\mu$) 
\begin{equation}
\label{J_4_spinor_field}
{\cal J }
 = 
 {\scriptsize
\begin{pmatrix}
      -J_x + i J_y    \\
      \rho +J_z\\
    -\rho +J_z\\
     J_x + i J_y  \\
\end{pmatrix}
}.
\end{equation}
Therefore, (\ref{Maxwell_equations_spinor_form})   encodes the equations of motion
 \begin{eqnarray}
i \partial_t \bm{F} & = & 
\nabla\times\bm{F} - i e \bm{J}
\\
\nabla\cdot \bm{F} & = & e \rho, 
\end{eqnarray}
 which are the Maxwell equations in 3-vector form for the complex electromagnetic field $\bm{F}$.

 \subsection{Coupled Weyl equations}

   Using the  4-spinor fields in (\ref{QLG_algorithm_for_the_Maxwell_field_block_diagonal_four_spinor_form}) and (\ref{J_4_spinor_field}), 
    Maxwell's equation for the photon field ${\cal A}$, electromagnetic field ${\cal F}$, and source field  ${\cal J}$,   respectively
\begin{equation}
{\cal A} 
 =  
 {\scriptsize
\begin{pmatrix}
      -A_x + i A_y    \\
  A_0+   A_z \\
   -A_0+   A_z\\
     A_x + i A_y  \\
\end{pmatrix}
}
\quad
{\cal F}
 =  
 {\scriptsize
\begin{pmatrix}
     - F_x + i F_y    \\
       - \partial\cdot A  + F_z \\
        \partial\cdot A + F_z \\
      F_x + i F_y  
\end{pmatrix}
}
\quad
{\cal J} 
 = 
 {\scriptsize
\begin{pmatrix}
      -J_x + i J_y    \\
   \rho  +J_z\\
    -\rho +J_z\\
     J_x + i J_y  \\
\end{pmatrix}
},
\end{equation}
are the coupled quantum wave equations
\begin{subequations}
\begin{eqnarray}
\left(-\partial_t +\bm{1}\otimes \bm{\sigma}\cdot \nabla\right)
{\scriptsize
\begin{pmatrix}
      -A_x + i A_y    \\
  A_0+   A_z \\
   -A_0+   A_z\\
     A_x + i A_y  \\
\end{pmatrix}
}
&=&
{\scriptsize 
\begin{pmatrix}
     - F_x + i F_y    \\
       - \partial\cdot A  + F_z \\
        \partial\cdot A + F_z \\
      F_x + i F_y  
\end{pmatrix}
}
\\
\left(-\partial_t -\bm{1}\otimes \bm{\sigma}\cdot \nabla\right)
{\scriptsize 
\begin{pmatrix}
     - F_x + i F_y    \\
       - \partial\cdot A  + F_z \\
        \partial\cdot A + F_z \\
      F_x + i F_y  
\end{pmatrix}
}
&=&
e
{\scriptsize 
\begin{pmatrix}
     - J_x + i J_y    \\
   \rho  +J_z\\
   - \rho +J_z\\
     J_x +i J_y  \\
\end{pmatrix}
}.
\end{eqnarray}
\end{subequations}
Moreover, in tensor-product form, Maxwell's equations are fully specified by the quantum wave equations
\begin{subequations}
\begin{eqnarray}
\label{Maxwell_equation_4spinor_rep}
{\cal F} &=&
 (-\partial_t 
+ \bm{1} \otimes\bm{\sigma}\cdot \nabla){\cal A}, 
\\
e {\cal J} &=&
 (-\partial_t 
- \bm{1}\otimes \bm{\sigma}\cdot \nabla) {\cal F}, 
\end{eqnarray}
\end{subequations}
taking the 4-divergence to be zero (Lorentz gauge). Using the definitions
\begin{equation}
\sigma^\mu = (1,\bm{\sigma}) \qquad  \bar\sigma^\mu=(1,-\bm{\sigma}),
\end{equation}
one can write Maxwell's equations in covariant form
\begin{subequations}
\label{Maxwell_equation_4spinor_rep_covariant_form_Derived}
\begin{eqnarray}
-{\cal F} &=&
 \bm{1} \otimes  \bar\sigma\cdot \partial {\cal A}, 
\\
-e {\cal J} &=&
 \bm{1} \otimes\sigma\cdot \partial {\cal F}, 
\end{eqnarray}
\end{subequations}
using the shorthand notation $\sigma\cdot \partial = \sigma^\mu \partial_\mu$ and $\bar\sigma\cdot \partial = \bar\sigma^\mu \partial_\mu$.  
These are  the Maxwell equations  (\ref{Maxwell_equation_4spinor_rep_covariant_form_a}) and (\ref{Maxwell_equation_4spinor_rep_covariant_form_b}).

\section{Derivation the quantum equation for  4-spinor  fields ${\cal A}$ and $\tilde {\cal A}$}
\label{Sec_derivation_Maxwell_London_equations_Yepez_form}

With London penetration depth $\lambda_\text{\tiny $L$}$,  the superconducting ansatz is
\begin{equation}
e {\cal J} =-\frac{1}{\lambda_\text{\tiny $L$}^2} {\cal A}
\end{equation}
and the Maxwell-London equations are
\begin{subequations}
\begin{eqnarray}
-{\cal F} &=&
 \bm{1} \otimes \bar\sigma\cdot \partial {\cal A}, 
\\
\frac{1}{\lambda_\text{\tiny $L$}^2} {\cal A}&=&
 \bm{1} \otimes\sigma\cdot \partial {\cal F}.
\end{eqnarray}
\end{subequations}
If we multiply the first equation through by $i$ and the second equation through by $\lambda_\text{\tiny $L$}$, then the equations of motion may be written as 
\begin{subequations}
\begin{eqnarray}
\label{Maxwell_equation_4spinor_rep_form2}
\left.
\frac{1}{\lambda_\text{\tiny $L$}} 
\middle(
 -i \lambda_\text{\tiny $L$}  {\cal F}
 \right)
  &=&
i\bm{1} \otimes\bar\sigma\cdot \partial {\cal A}, 
\\
\frac{1}{\lambda_\text{\tiny $L$}} {\cal A}&=&
i \bm{1}\otimes \sigma\cdot \partial (-i\lambda_\text{\tiny $L$}  {\cal F}) .
\end{eqnarray}
\end{subequations}
Furthermore, if we define a  dual
 Maxwell 4-spinor
\begin{equation}
\tilde {\cal A} \equiv -i\lambda_\text{\tiny $L$}  {\cal F}
,
\end{equation}
then  the equations of motion take the symmetrical form
\begin{subequations}
\begin{eqnarray}
\label{Maxwell_equation_4spinor_rep_form3}
\frac{1}{\lambda_\text{\tiny $L$}} \tilde {\cal A}&=&
i\bm{1} \otimes\bar\sigma\cdot \partial {\cal A}, 
\\
\frac{1}{\lambda_\text{\tiny $L$}} {\cal A}&=&
i \bm{1} \otimes\sigma\cdot \partial \tilde {\cal A}. 
\end{eqnarray}
\end{subequations}
Then, we can write these equations of motion as a single quantum wave equation in matrix form
\begin{equation}
\begin{pmatrix}
  - \frac{1}{\lambda_\text{\tiny $L$}}     &  i \bm{1}\otimes\sigma \cdot \partial  \\
  i \bm{1}\otimes\bar{\sigma} \cdot \partial    & - \frac{1}{\lambda_\text{\tiny $L$}} 
\end{pmatrix}
\begin{pmatrix}
      {\cal A}    \\
     \tilde {\cal A}
\end{pmatrix}
= 0.
\end{equation}
Taking $\lambda_\text{\tiny $L$}$ to be 
\begin{equation}
\lambda_\text{\tiny $L$} = \frac{1}{m_\text{\tiny $L$}},
\end{equation}
where $m>0$ is real-valued, then the equations of motion are
\begin{equation}
\label{Yepez_equation}
\begin{pmatrix}
  - m_\text{\tiny $L$}    &  i \bm{1}\otimes\sigma \cdot \partial  \\
  i \bm{1}\otimes\bar{\sigma} \cdot \partial    & - m_\text{\tiny $L$}
\end{pmatrix}
\begin{pmatrix}
      {\cal A}    \\
     \tilde {\cal A}
\end{pmatrix}
= 0.
\end{equation}
The form of the Maxwell-London equations (\ref{Yepez_equation})  is equivalent to (\ref{Maxwell_London_equations_1}) and (\ref{Maxwell_London_equations_2}).

\section{Derivation paired 4-spinor quantum equation for the field $\Phi$}
\label{Sec_derivation_Maxwell_London_equations_Yepez_generalized_Dirac_form}

Here  the form of the equation of motion (\ref{Maxwell_London_equations_symmetrical_4_vector_form}) is established by showing it is equivalent to (\ref{Maxwell_London_equations_in_Yepez_form}) that was derived in the previous Appendix~\ref{Sec_derivation_Maxwell_London_equations_Yepez_form}.   The generalized Dirac equation for a pair of  4-spinors is
\begin{equation}
\label{4_spinor_pair_quantum_wave_equation}
i\hbar c \,{\cal G}_\mu \partial^\mu  \Phi
 =
    m_\text{\tiny $L$}c^2\Phi,
\end{equation}
for field
\begin{equation}
\Phi
\equiv
\begin{pmatrix}
      {\cal A}    \\
     \tilde {\cal A}
\end{pmatrix},
\end{equation}
where the generalized Dirac matrices are
\begin{equation}
{\cal G}_0 = \sigma_x\otimes \bm{1} \otimes\bm{1}
=
\begin{pmatrix}
0      &    \bm{1}\otimes \bm{1} \\
    \bm{1}\otimes \bm{1}   &  0 
\end{pmatrix}
\qquad
\bm{{\cal G}} = i\sigma_y\otimes \bm{1} \otimes\bm{\sigma}
=
\begin{pmatrix}
0      &    \bm{1}\otimes \bm{\sigma} \\
    \bm{1}\otimes \bm{\sigma}   &  0 
\end{pmatrix}.
\end{equation}
Expanding (\ref{4_spinor_pair_quantum_wave_equation}), and using natural units $\hbar=1$ and $c=1$, we have
\begin{equation}
i(  {\cal G}_0 \partial_0 +  \bm{{\cal G}}\cdot \nabla) \Psi -  m_\text{\tiny $L$}\Phi =0 .
\end{equation}
Multiplying through by ${\cal G}_0$ gives
\begin{equation}
i(  \partial_0 +   {\cal G}_0  \bm{{\cal G}}\cdot \nabla) \Psi -  m_\text{\tiny $L$}{\cal G}_0 \Phi =0 .
\end{equation}
Since
\begin{equation}
 {\cal G}_0  \bm{{\cal G}} = - \sigma_z \otimes\bm{1}\otimes\bm{\sigma}
 =
 \begin{pmatrix}
-\bm{1}\otimes\bm{\sigma}      & 0   \\
 0     &   \bm{1}\otimes\bm{\sigma}
\end{pmatrix},
\end{equation}
the quantum wave equation becomes
\begin{equation}
i\left[  \partial_0 +    \begin{pmatrix}
-\bm{1}\otimes\bm{\sigma}      & 0   \\
 0     &   \bm{1}\otimes\bm{\sigma}
\end{pmatrix}
\cdot \nabla
\right]
 \begin{pmatrix}
      {\cal A}    \\
     \tilde {\cal A}
\end{pmatrix}-  m_\text{\tiny $L$}
\begin{pmatrix}
0      &    \bm{1}\otimes \bm{1} \\
    \bm{1}\otimes \bm{1}   &  0 
\end{pmatrix} 
\begin{pmatrix}
      {\cal A}    \\
     \tilde {\cal A}
\end{pmatrix}
 =0 .
\end{equation}
This is the coupled set of equations
\begin{subequations}
\begin{eqnarray}
i (\partial_0 - \bm{1}\otimes\bm{\sigma}\cdot \nabla) {\cal A}  - m_\text{\tiny $L$} \tilde {\cal A} & = & 0 \\
i (\partial_0 + \bm{1}\otimes\bm{\sigma}\cdot \nabla)\tilde {\cal A}  - m_\text{\tiny $L$}  {\cal A} & = & 0 .
\end{eqnarray}
\end{subequations}
Finally, writing $\sigma \cdot \partial = \partial_0 + \bm{\sigma}\cdot\nabla$  and $\overline{\sigma} \cdot \partial = \partial_0 - \bm{\sigma}\cdot\nabla$, the quantum wave equation equation  in matrix form is
\begin{equation}
\label{Yepez_equation_Rederived}
\begin{pmatrix}
  - m_\text{\tiny $L$}    &  i \bm{1}\otimes\sigma \cdot \partial  \\
  i \bm{1}\otimes\bar{\sigma} \cdot \partial    & - m_\text{\tiny $L$}
\end{pmatrix}
\begin{pmatrix}
      {\cal A}    \\
     \tilde {\cal A}
\end{pmatrix}
= 0.
\end{equation}

\section{Bloch-Wannier continuous-field picture}
\label{Sec_Bloch_Wannier_continuous_field_picture}

In condensed matter theory,  quantum particle dynamics is confined to a spatial lattice, where the lattice that is  effectively produced by an external continuous periodic potential (due to a crystallographic arrangement of positively-charged atomic nuclei) of the form
\begin{equation}
\label{single_lattice}
V_\text{crys.}(x) = 
V_\circ \sin^2(k_x x)\sin^2(k_y x)\sin^2(k_z z).
\end{equation}
That is,  each minima of a well in the periodic lattice represents a lattice point.  The external potential (\ref{single_lattice}) is an example of a spatial lattice specified by a single  wave vector $\bm{k} = (k_x, k_y, k_z)$.\footnote{In solid-state physics, it is common to have a more complicated spatial lattice specified by two or more wave vectors, but that is not needed for the modeling purposes here.}  It is conventional to employ delocalized periodic wave functions---a complete set of orthogonal energy eigenstates called called Bloch waves---to represent the  state of a quantum particle in  the lattice \cite{Bloch_1929}.   
 The energy eigenstates of 
 the system Hamiltonian may be written in the form
\begin{equation}
\label{crystal_energy_eigenstate}
\phi^{(n)}_{\bm{k}}(\bm{x}) = e^{i \bm{k} \cdot \bm{x}} u_{\bm{k}}^{(n)}(\bm{x}),
\end{equation}
where $u_{\bm{k}}^{(n)}(\bm{x})=u_{\bm{k}}^{(n)}(\bm{x}+\bm{a})$ denotes the  periodic Bloch wave for the crystal specified by wave vector $\bm{k}$ and lattice cell size $\bm{a}$.

Additionally, in the tight-binding approximation, it is  conventional to use localized (nonperiodic) states to represent a quantum particle in the lattice, where the quantum particle's wave function is narrow and  positioned  at the minimum of one of the wells of the periodic potential.  That is, the quantum particle's average position is centered at the well's minimum.  These localized wave packets are called Wannier functions \cite{PhysRev.52.191}, and they may be written as a superposition of Bloch waves (for simplicity say along the $x$-axis) as
\begin{equation}
w_n(x-x_i) = \frac{1}{2N} \sum_{\nu=-N}^{N-1} e^{-i k_\nu x_i} \phi_{k_\nu}^{(n)}(x),
\end{equation}
for $k_\nu = \pi \nu/(Na)$ with a crystal cellsize $a$.  If we assume the Bloch wave is independent of wave number and write the energy eigenstate as $\phi_{k_\nu}^{(n)}(x)=e^{i k_\nu x} u^{(0)}(x)$, then we may in turn write the Wannier functions as  
\begin{subequations}
\begin{eqnarray}
w_n(x-x_i) &=& \frac{u^{(0)}(x)}{2N} \sum_{\nu=-N}^{N-1} e^{i k_\nu (x-x_i)} 
\\
\label{Wannier_function_k_independent_Bloch_waves}
&=& 
\frac{u^{(0)}(x)}{2N} \sum_{\nu=-N}^{N-1} e^{\frac{\pi i \nu}{N a}(x-x_i)} .
\end{eqnarray}
\end{subequations}
We can make use of the finite geometric series identity 
\begin{equation}
\frac{1}{2N} \sum_{\nu=-N}^{N-1} z^\nu = \frac{1}{2N}\frac{z^N-z^{-N}}{z-1}
\end{equation}
to rewrite (\ref{Wannier_function_k_independent_Bloch_waves}) as 
\begin{subequations}
\begin{eqnarray}
\nonumber
w_n(x-x_i) &=&
\frac{ u^{(0)}(x)}{2N} \frac{e^{\frac{\pi i }{ a}(x-x_i)}-e^{-\frac{\pi i }{a}(x-x_i)}}{e^{\frac{\pi i }{N a}(x-x_i)}-1}
\\
\\
&=& 
\frac{i u^{(0)}(x)}{N}\frac{\sin(\pi(x-x_i)/a)}{e^{\frac{\pi i }{N a}(x-x_i)}-1}
\\
&=& 
{ u^{(0)}(x)}\frac{\sin(\pi(x-x_i)/a)}{\pi  (x-x_i)/a} + \cdots
\qquad
\quad
\\
&\approx&
{ u^{(0)}(x)}\, {\text{sinc}(\pi(x-x_i)/a)},
\end{eqnarray}
\end{subequations}
which is indeed a localized wave packet centered at $x_i$. 
Both the Bloch wave and Wannier function representations have the advantage of allowing one to treat a lattice-based condensed matter system with continuous probability amplitude fields defined in a continuous space.

\section{Streaming protocol}
\label{Streaming_protocol}

 Consider the matter field $\psi$.  Suppose for the moment (for the sake of pedagogy) that the unitary operator ${\cal C}(x)$ is represented by  an
 identity matrix (so there is no unitary mixing of the left and right components of $\psi$ for $m=0$). Then, 
 we may rewrite (\ref{Simplest_QLG_model}) simply as the update rule
%
%
\begin{equation}
\label{Simplest_QLG_model_chiral}
\psi(x-  \ell \gamma(x)  - i \ell \gamma_0 \cdot \gamma(x) \cdot e A(x)/(\hbar c)) = \psi(x ) .
\end{equation}
So, in the Bloch-Wannier picture,  (\ref{Simplest_QLG_model_chiral}) can be written  in exponential form
\begin{equation}
\label{Simplest_QLG_model_chiral_exponential_form}
e^{-\ell (\gamma^\mu(x) \partial_\mu- i  \gamma_0 \cdot \gamma^\mu(x)  e A_\mu(x)/(\hbar c))} \psi(x)=\psi(x),
\end{equation}
where the covariant derivative is $\partial_\mu = ( \partial_t, \nabla)$. 
Equation of motion (\ref{Simplest_QLG_model_chiral_exponential_form}) is an exact representation of particle dynamics confined to a spacetime lattice because the operator   ${\cal S}^\dagger(x) =e^{-\ell (e^\mu(x) \partial_\mu- i  e_0 e^\mu(x)  e A_\mu(x)/(\hbar c))}$ is actually unitary and causes a particle to hop between neighboring points $x^\mu \rightarrow x^\mu - \ell e^\mu(x)$ as well as to undergo an 
 unitary gauge transformation along the way.  The evolution equation  (\ref{Simplest_QLG_model_chiral_exponential_form}) is well defined in the sense that if a particle  exists at a point on the lattice before the application of ${\cal S}^\dagger $, then that particle will  exist at a point on the lattice after its application, and in general  the particle will become quantum mechanically entangled with other particles at the same point.

For $A^\mu=0$, (\ref{Simplest_QLG_model_chiral}) reduces to
\begin{equation}
\label{effective_chiral_QLG_EoM_2}
-\ell \gamma^\mu\partial_\mu \psi(x) + \cdots = 0.
\end{equation}
It is conventional to identify the equation of motion for a chiral $\psi$ field as the Euler-Lagrange equation that derives from applying the least action principal  $\delta S=0$, where the action is $S =\int dx^4 {\cal L}$ and where ${\cal L} = i\hbar c \overline{\psi} \gamma^\mu \partial_\mu \psi$ is the covariant Lagrangian density that has units of energy density.  To  conform to this convention, one is free to multiply the equation of motion (\ref{effective_chiral_QLG_EoM_2}) by a highest-energy scale (or Planck scale energy) in the quantum lattice gas model that is taken to be $\hbar/\tau$.   Then, (\ref{effective_chiral_QLG_EoM_2})   becomes the Weyl equation in Minkowski space
\begin{equation}
\label{Weyl_equation_approximation}
i \hbar c \gamma^\mu \partial_\mu \psi(x) + \cdots = 0,
\end{equation}
where it is conventional to multiply by the imaginary number as well. 

The 4-spinor field has left- and right-handed 2-spinor components
\begin{equation}
\label{Dirac_4_spinor}
\psi(x)=
\begin{pmatrix}
      \psi_\text{\tiny L}(x)
          \\
      \psi_\text{\tiny R}(x)
\end{pmatrix}
=
{\scriptsize
\begin{pmatrix}
      \psi_{\text{\tiny L}\uparrow}(x)
      \\
      \psi_{\text{\tiny L}\downarrow}(x)
      \\
      \psi_{\text{\tiny R}\uparrow}(x)
      \\
      \psi_{\text{\tiny R}\downarrow}(x)
      \\
\end{pmatrix}
},
\end{equation}
so in the chiral representation $\gamma^\mu= (\gamma^0, \bm{\gamma})=( \sigma_x \otimes\bm{1} , i\sigma_y\otimes \bm{\sigma})$   the lefthand side of (\ref{Weyl_equation_approximation}) may be written as
\begin{subequations}
\begin{eqnarray}
i \hbar c \gamma^\mu \partial_\mu \psi(x) 
&=& 
i\hbar c \left(\gamma^0 \partial_t +\bm{\gamma} \cdot \nabla \right)\psi(x)
\\
&=& 
i\hbar c \left( \sigma_x \otimes\bm{1} \partial_t+  i\sigma_y \otimes\bm{\sigma}\cdot \nabla \right)\psi(x)
\qquad
\\
\nonumber
&=& 
i\hbar c 
\begin{pmatrix}
    0  &   \partial_t +  \bm{\sigma}\cdot \nabla  \\
\partial_t -  \bm{\sigma}\cdot \nabla      &  0
\end{pmatrix}
\begin{pmatrix}
      \psi_\text{\tiny L}(x)
          \\
      \psi_\text{\tiny R}(x)
\end{pmatrix},
\\
\end{eqnarray}
\end{subequations}
where the Pauli spin vector is $\bm{\sigma} = (\sigma_x, \sigma_y, \sigma_z)$. 
Then, the equation of motion (\ref{Weyl_equation_approximation}) reduces to independent equations of motion
\begin{subequations}
\label{chirality_separated_Weyl_equations}
\begin{eqnarray}
( \partial_t +  \bm{\sigma}\cdot \nabla) \psi_\text{\tiny R}(x) &=&0
\\
( \partial_t -  \bm{\sigma}\cdot \nabla) \psi_\text{\tiny L}(x) &=&0,
\end{eqnarray}
\end{subequations}
where the left- and right-handed 2-spinors fields move in opposite directions.

The  quantum lattice gas algorithm,  without any interactions, is based on the unitary representation of (\ref{chirality_separated_Weyl_equations})
\begin{subequations}
\label{chirality_separated_Weyl_equations_unitary_form}
\begin{eqnarray}
 \psi'_\text{\tiny R}(x) &=&  e^{\ell \bm{\sigma}\cdot \nabla} \psi_\text{\tiny R}(x) 
\\
 \psi'_\text{\tiny L}(x) &=&  e^{-\ell \bm{\sigma}\cdot \nabla} \psi_\text{\tiny L}(x),
\end{eqnarray}
\end{subequations}
which is equivalent to (\ref{Simplest_QLG_model_chiral_exponential_form}) when the gauge field induced phase rotation  of $\psi$ is added.  For example, the simplest quantum lattice gas algorithm one can write for (\ref{chirality_separated_Weyl_equations_unitary_form}) is
\begin{subequations}
\label{chirality_separated_Weyl_equations_unitary_form_quantum_algorithm}
\begin{eqnarray}
 \psi'_\text{\tiny R}(x) &\cong&  e^{\ell \sigma_x\partial_x}e^{\ell \sigma_y\partial_y}e^{\ell \sigma_z\partial_z} \psi_\text{\tiny R}(x) 
\\
 \psi'_\text{\tiny L}(x) &\cong& e^{-\ell \sigma_x\partial_x}e^{-\ell \sigma_y\partial_y}e^{-\ell \sigma_z\partial_z} \psi_\text{\tiny L}(x).
\end{eqnarray}
\end{subequations}

\acknowledgments 
 
 I would like to thank Norman Margolus and Xerxes Tata for reviewing this manuscript and providing helpful comments.  
 Quantum lattice models have developed from the Air Force Quantum Computing Program that began in the mid 1990's. 
This work was supported by a grant  ``Quantum information dynamics intrinsic to strongly-correlated Fermi systems"  from the Air Force Office of Scientific Research. 



\end{document}